\documentclass[11pt]{article}
\usepackage{graphicx}
\usepackage{mathptmx, mathptmx, amsfonts, amsmath}

\usepackage[unicode]{hyperref}
\hypersetup{pdftitle=RESS Paper,
pdfauthor={Kenbeek, Kapodistria, Di Bucchianico},
pdftitle={Data-driven online  monitoring of wind turbines},
pdfstartview=FitH,
pdfpagemode=UseOutlines,
hidelinks
}
\usepackage{lineno}
\usepackage{siunitx}
\modulolinenumbers[5]

\usepackage{graphicx}
\usepackage{caption}
\usepackage{subcaption}
\DeclareGraphicsExtensions{.png,.pdf}

\usepackage{xy}

\usepackage{color, colortbl}
\definecolor{Gray}{gray}{0.9}

\usepackage[square,sort,comma,numbers]{natbib}

\newcommand{\sep}{; }
\newcommand{\MSC}{{\em MSC 2010:}}

\title{Data-driven online  monitoring of wind turbines}

\author{Thomas Kenbeek $\S$ \and Stella Kapodistria \footnote{Corresponding author email:    \texttt{s.kapodistria@tue.nl}}\ \ $\S$ \and Alessandro Di Bucchianico $\S$}
\footnotetext{$\S$ Department of Mathematics and Computer Science, Eindhoven University of Technology, The~Netherlands,}
\date{November, 2016}

\begin{document}

\maketitle

\begin{abstract}
Condition based maintenance is a modern approach to maintenance which has been successfully used in several industrial sectors. In this paper we present a concrete statistical approach to condition based maintenance  for wind turbine  by applying ideas from statistical process control. A specific problem in wind turbine maintenance is that failures of a certain part may have causes that originate in other parts a long time ago. This calls for methods that can produce timely warnings by combining sensor data from different sources. Our method improves on existing methods used in wind turbine maintenance by using adaptive alarm thresholds for the monitored parameters that correct for values of other relevant parameters. We illustrate our method with a case study that shows that our method is able to  predict upcoming failures much earlier than currently used methods.\\

\MSC  ~90B25\sep  62P30\sep 62J05\sep 62M10\sep 62L10
\end{abstract}

\section{Introduction}
Condition based maintenance is a modern approach to predictive maintenance that aims at optimizing maintenance decisions by using information obtained from monitoring the condition of the equipment to be maintained rather than information obtained from scheduled maintenance activities. Recent advances in sensor technology make this approach very attractive, since it has become easier (both technologically and financially) to obtain up-to-date relevant monitoring data.

Maintenance of wind turbines is very important since damage to wind turbines may be very costly and lead to long periods of unavailability due to the time-consuming repairs. This is in particular true for offshore wind turbines. An important reason for the relevance of applying condition based maintenance to wind turbines rather than periodic maintenance is that failures in wind turbines may be caused by damages that occurred much earlier and at an other location in the turbine, e.g., cracks in a rotor blade caused by lightning may propagate to the generator or the gearbox.

In order to successfully apply condition based maintenance, it is thus important to have a method for delivering  timely warnings about changes in the condition of a wind turbine. Current practice is to base warnings on fixed ISO thresholds, especially for vibration readings. However, we need to note that temperatures do not have such thresholds and even for vibration readings, this is not a correct approach since it ignores the relation between several parameters, e.g., the vibration readings are influenced by the wind speed. So in order to judge whether the underlying reading is abnormal or not, we have to take into account external factors. Instead, we propose to use regression analysis to adjust sensor values of parameters in order to obtain warnings for failures based on  abnormal sensor values of properly adjusted relevant parameters. The adjusted values are then monitored using control charts, the standard tool in SPC (Statistical Process Control). The regression models may be either black-box models or white-box models based on physical laws.

\section{Problem Description}\label{sec:problemdescription}
In this paper we analyze wind turbine  data provided to us by a commercial maintenance company in collaboration with a wind park owner (an electricity company).
 The objective is the creation of models which permit the scheduling of predictive maintenance of wind
 turbines by predicting failures in a timely manner. To this purpose, we analyzed a large amount of data, consisting of
 condition data, operational data and environmental data, together with maintenance and event logs, that have been collected over a period of almost two years. Our models use all provided data.

\section{ Technical Background}\label{sec:background}
Many studies have been conducted to evaluate the reliability of wind turbines as the integrated part of the grid, see, e.g., \cite{hu2009reliability,kahrobaee2010short,lingling2010reliability}. Some of these studies have addressed the individual wind turbine reliability modeling, and investigated the major factors contributing to the total failure of the turbine, see, e.g., \cite{SpinatoTavnerBusselEtAl2009,guo2009reliability}. Reliability data about wind turbine assemblies have become available in recent years from surveys, see, e.g., \cite{SpinatoTavnerBusselEtAl2009,tavner2007reliability}. Table~\ref{tab:FMEA1} shows a typical comparison between reliability field data of a small wind turbine, 300kW, and a 1MW wind turbine main assembly failure rates based on \cite{SpinatoTavnerBusselEtAl2009}.

\begin{table}[htb]
\centering
\begin{tabular}{|l | l | l|}
\hline
Assembly & \multicolumn{2}{r|}{\txt{Failure rate of LKW WTs \\(Failure per turbine per year)}} \\
\cline{2-3}
  & 300 kW WT & 1MW WT \\
\hline
Generator & 0.059 & 0.126 \\
Brake & 0.029 & 0.056 \\
Hydraulics & 0.039 & 0.096 \\
Yaw system & 0.079 & 0.152 \\
Sensors & 0.037 & 0.151 \\
Pitch system & 0.034 & 0.237\\
Blade & 0.078 & 0.308 \\
Gearbox & 0.079 & 0.255 \\
Shaft/bearings & 0.002 & 0.046 \\
\hline
\end{tabular}
\caption{Wind turbine assemblies' reliability field data}
\label{tab:FMEA1}
\end{table}

Besides reliability analysis, Failure Mode and Effects Analysis (FMEA) is performed in order to determine several key potential failures in the system through the comparison of some predefined factors. As  a result, such an analysis helps increase the availability of that system, see, e.g., \cite{stamatis2003failure,mikulak2008basics}. This process has been used on almost any equipment from cars to space shuttles, and as of the last decade wind turbines have been subjected to such analyses, e.g., \cite{klein1990model,andrawus2006selection,tavner2010using}.

\subsection{Failure modes}\label{subsec:failuremodes}
Failures are defined as situations in which a device no longer operates the way intended. There are numerous failure modes that can be defined for a complicated assembly like a wind turbine. These failure modes may cause partial or complete loss of power generation. Mainly, the key failure modes, which cause complete loss of power generation, are malfunction and major damage of the main parts of the turbine. Other failure modes are less significant and include surface damage and cracks, oil leakage, loose connection, etc. However, if they are not taken care of, minor failure modes can initiate major failures as well, see for more details Table~\ref{tab:FMEA2} based on \cite{kahrobaee2011risk}. Evidently, every failure mode has a root cause, and the probability of that failure mode is directly  related to the probability of its root cause. Table~\ref{tab:FMEA3}, based on \cite{kahrobaee2011risk}, provides different categories for these causes. Human error in this table, refers to the errors occurring during operation or maintenance.

\begin{table}[htb]
\centering
\begin{tabular}{|l | l|}
\hline
Sub-assemblies & Main Parts \\
\hline
Structure &  Nacelle, Tower, Foundation \\
Rotor  & Blades, Hub, Air brake \\
Mechanical Brake  & Brake disk, Spring, Motor \\
Main shaft &  Shaft, Bearings, Couplings \\
Gearbox &  Toothed gear wheels, Pump,Oil heater/cooler, Hoses \\
Generator  & Shaft, Bearings, Rotor, Stator, Coil \\
Yaw system &  Yaw drive, Yaw motor \\
Converter  & Power electronic switch, cable, DC bus \\
Hydraulics &  Pistons, Cylinders, Hoses \\
Electrical System &  Soft starter, Capacitor bank, Transformer, Cable, Switch gear \\
Pitch System &  Pitch motor, Gears \\
Control system  & Sensors, Anemometer, Communication parts, Processor, Relays \\
\hline
\end{tabular}
\caption{General set of wind turbine assemblies and main parts}
\label{tab:FMEA2}
\end{table}

\begin{table}
\centering
\begin{tabular}{|l| l| l| l|}
\hline
Weather & Mechanical & Electrical & Wear \\
\hline
High wind & Manufacturing and material defect & Grid fault & Ageing \\
Icing & Human error & Overload & Corrosion \\
Lightning & External damage & Human error & \\
 & External damage & Software failure & \\
\hline
\end{tabular}
\caption{Root causes of failure modes}
\label{tab:FMEA3}
\end{table}

\subsection{Failure detection}\label{subsec:failuredetection}
There are a variety of ways to detect the probable failure modes as categorized in Table~\ref{tab:FMEA4} based on \cite{kahrobaee2011risk}. The common ways are through inspection or while the turbine is being maintained. However, the fastest and the most reliable method is condition monitoring which can increase the availability of the wind turbine considerably by using on-line systems. With condition monitoring, the probability of not detecting the failure decreases to the failure probability of the human error or the monitoring system itself. The objective of this paper is to propose a statistical approach that can be used to timely detect  failures.

\begin{table}[htb]
\centering
\begin{tabular}{|l | l | l|}
\hline
Inspection & Condition Monitoring & Maintenance \\
\hline
Visual & Vibration analysis & Time-Based \\
Olfactive & Oil analysis & Condition-Based \\
Auditive & Infrared thermography & \\
 & Ultrasonic & \\
\hline
\end{tabular}
\caption{Major detection methods of the failure modes}
\label{tab:FMEA4}
\end{table}

\section{Data Description}\label{sec:data}
The data was collected from a Vestas V47 wind turbine built in 2000. The height of the turbine is $\SI{65}{\metre}$, the rotor diameter is $\SI{47}{\metre}$ and the rotor sweep is $\SI{17.35}{\metre \squared}$. The typical power output for the Vestas V47 turbine is $\SI{600}{\kW}$. It joins the grid connection at a wind speed of $\SI{4}{\metre \per \second}$ with a rated actual power output (typically achieved) at a wind speed of $\SI{15}{\metre \per \second}$, and it is disconnected at a wind speed of $\SI{25}{\metre \per \second}$. Furthermore, the Vestas V47 turbine is designed to function up to a maximum wind speed of $\SI{59.5}{\metre \per \second}$.

Typically, the Vestas V47 turbine is equipped with a single generator. However, the data comes from a turbine equipped with two generators. The second generator is smaller and is only used when the wind speed is very low (less than $\SI{7}{\metre \per \second}$). The main benefits of a second generator are a lower sound level and an increased power production at low wind speeds.

\begin{figure}[htb]
\centering
\includegraphics[scale=0.37]{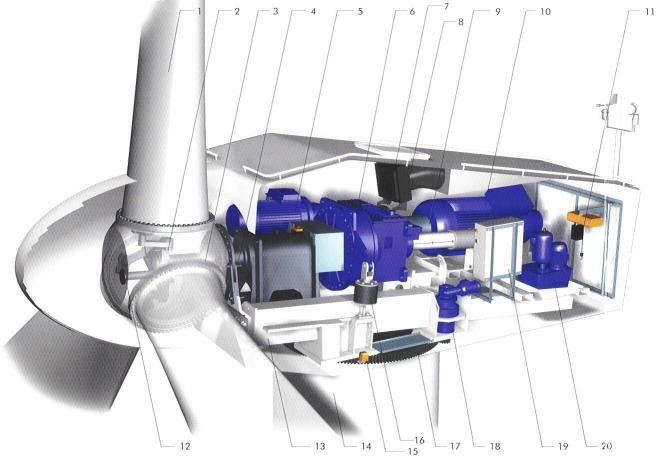}
\caption{The interior of the V47 nacelle. All parts are listed in Table \ref{tab:partstable}}
\label{img:schematics}
\end{figure}
\begin{table}[htb]
\centering
\begin{tabular}{|r l| r l|}
\hline
1.& Blade & 11. & Service crane \\
2.& Blade hub & 12. & Pitch cylinder \\
3.& Blade bearing & 13. & Machine foundation \\
4.& Main shaft & 14. & Tower \\
5.& Secondary generator & 15. & Yaw control \\
6.& Gearbox & 16. & Gear tie rod \\
7.& Disc break & 17. & Yaw ring \\
8.& Oil cooler & 18. & Yaw gears \\
9.& Cardan shaft & 19. & VMP top control unit \\
10.& Primary generator & 20. & Hydraulic unit \\
\hline
\end{tabular}
\caption{The different parts of the Vestas V47 wind turbine}
\label{tab:partstable}
\end{table}

The turbine consists of several parts. These parts are listed in Table~\ref{tab:partstable} and a visualization of the Vestas V47 turbine is provided in Figure~\ref{img:schematics} taken from the Vestas V47 brochure \cite{Vestas-brochure}.
On December 2014, the wind turbine under consideration suffered a mechanical failure of the primary generator. While the domain experts suspected the problem, the identification of the root cause and the extend of the problem became apparent only after the failure. Our objective is to check whether the data could have assisted in identifying a change in the pattern and whether we could only from the data predict the failures well in advance. In this sense, one of the key objectives of this manuscript is to develop predictive models based on data so as to predict imminent failures, and identify the root cause. Finally, it is worth mentioning that after the failure of the generator, several other components such as the bearing and gearbox exhibited stress and failed (or were preventive replaced) soon after the generator failure.


For the development of the statistical approaches with the objective of detecting imminent failures, we use a data set containing $326160$ observations on $110$ variables. The data were collected in the period from $\text{19/06/2013}$ to $ \text{18/03/2015}$. In addition to the data set, we use the maintenance and event logs for the period of interest. In particular, all information we use for the development of the statistical approaches can be clustered into three categories:
\begin{enumerate}
\item \textbf{Environmental variables}: These describe the environmental factors. In our data set, the environmental variables are the wind speed and the environmental temperature.
\item \textbf{Conditional variables}: These variables describe the state of several parts of the turbine. The conditional variables can be further categorized into
\begin{enumerate}
\item \textbf{Speed}: This includes rotor speed and generator speed. Note that generator speed does not distinguish between the primary generator and the secondary generator.
\item \textbf{Temperatures}: This includes readings of the bearing, the gearbox, the two generators, the nacelle and the oil temperatures.
\item \textbf{Vibrations}: These include various vibration measurements from the gearbox input shaft,  the gearbox intermediate shaft, gearbox output shaft, the generator bearing drive end, the generator bearing non drive end, the planetary gearbox, the shaft bearing gearbox side, and the shaft bearing propeller side. There vibration readings include both acceleration and velocity measured in mm/s.

\item \textbf{Operational readings}: This includes the power output, the operating state, the pitch angle, and the yaw.
\end{enumerate}
\item \textbf{Maintenance and event logs}: Apart from the data set, we were also provided with access to the processed maintenance logs of the past two years. Any time any maintenance is performed on the turbine, the date and time is logged, as well as a short description of the work that was done, and if applicable the corresponding result, e.g. ``Ultrasonic scanning of blade bolts''.
\end{enumerate}
For the purpose of developing failure prediction models, we only use data collected during instances in which the turbine is running. To this end, we filter the data using the variable ``operating state''. This variable takes 4 values: 0 to indicate emergency, 1 to indicate stop, 2 to indicate pause and 3 to indicate run. We restrict our analysis to the cases the variable operating state takes value 3.

\section{Exploratory Data Analysis}\label{sec:EDA}
As a first step, so as to get a preliminary understanding of the data we present in this section plots of the individual variables versus time, as well as some scatter plots in order to visually identify patterns and relations between variables.

\begin{figure}
    \centering
    \begin{subfigure}[b]{0.45\textwidth}
        \includegraphics[width=\textwidth]{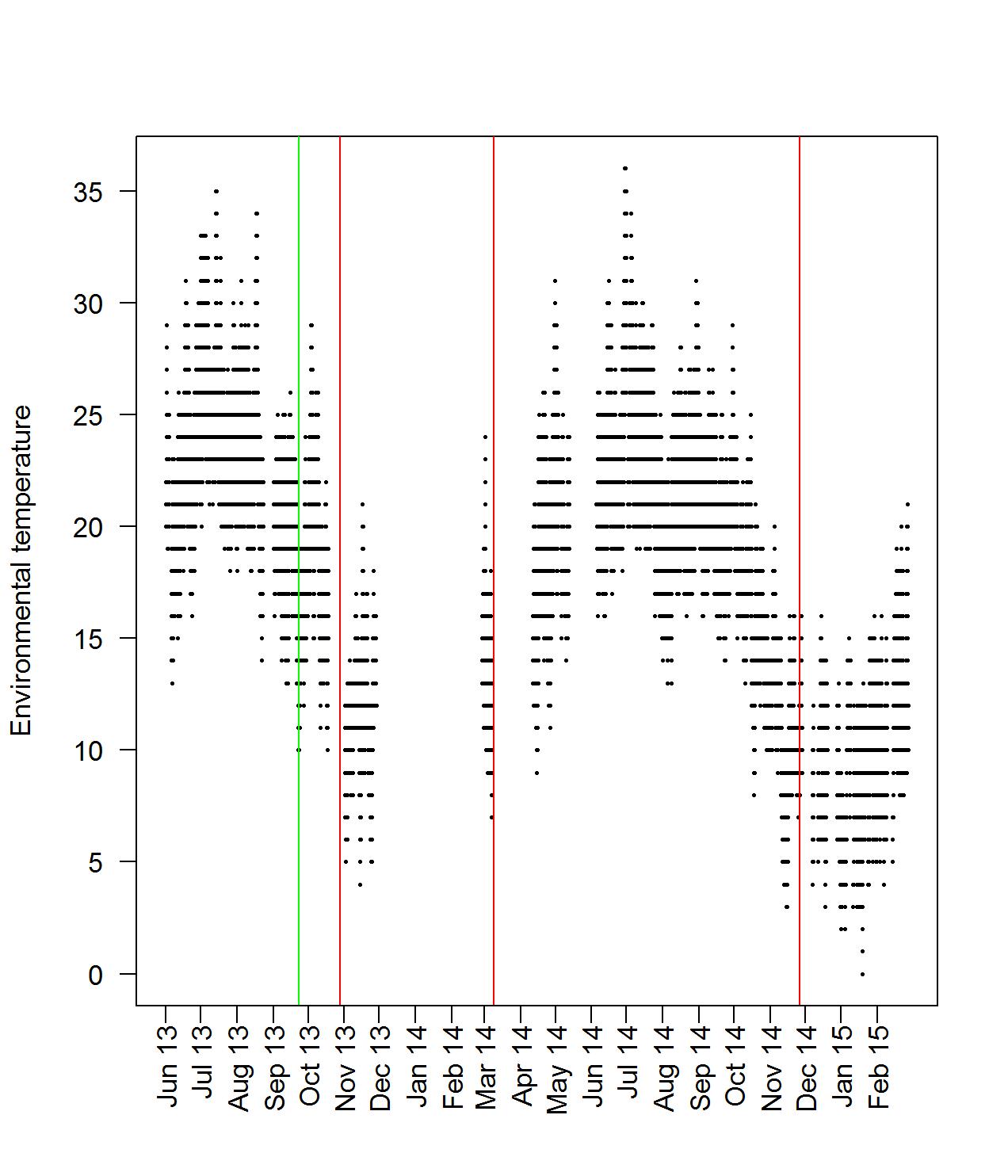}
        \caption{Environmental temperature}
        \label{fig:env_vs_time}
    \end{subfigure}
    ~
   \begin{subfigure}[b]{0.45\textwidth}
        \includegraphics[width=\textwidth]{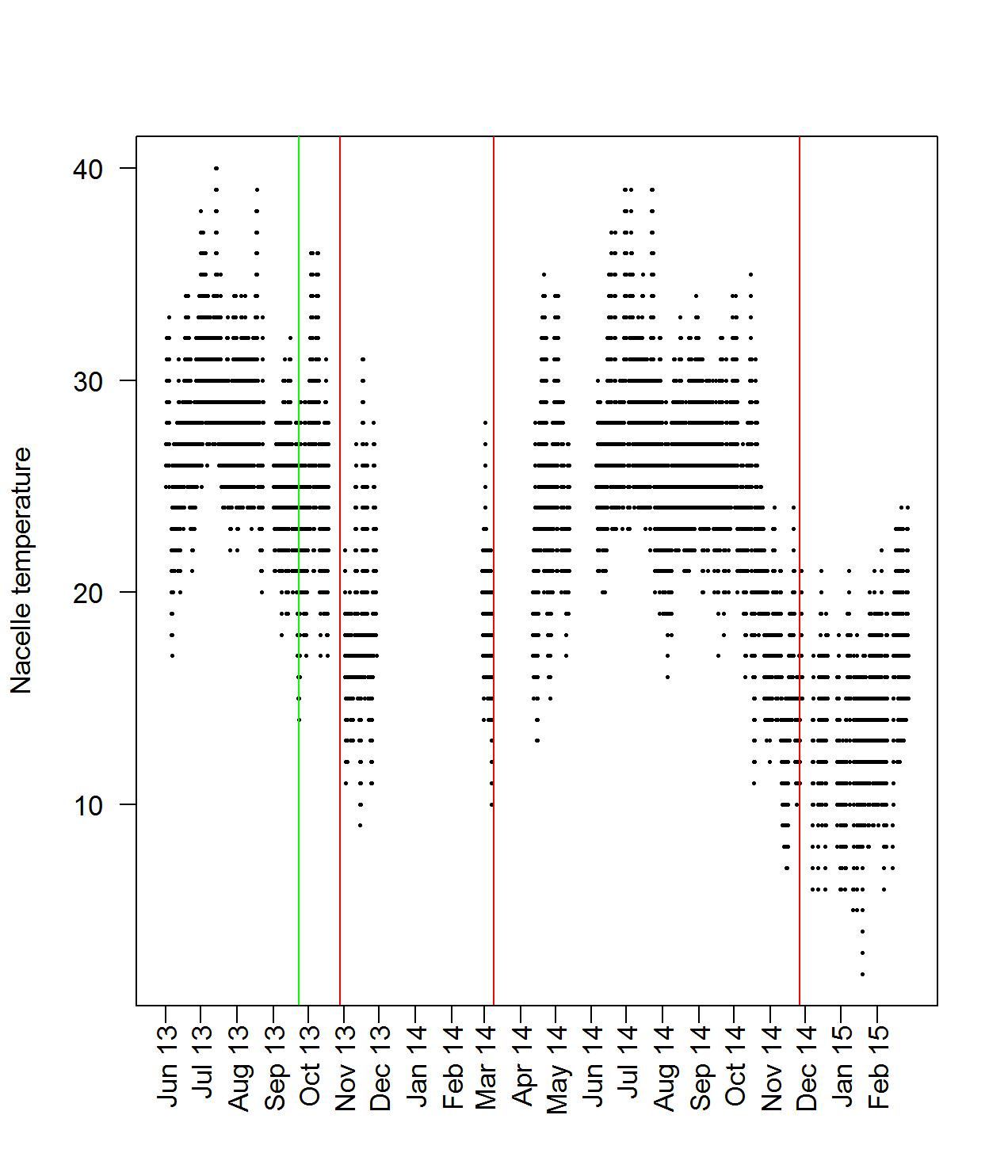}
        \caption{Nacelle temperature}
        \label{fig:nac_vs_time}
    \end{subfigure}
     \caption{The environmental and the nacelle temperature}
\label{fig:env_nacelle_oil_temp}
\end{figure}

First, we plot in Figure~\ref{fig:env_nacelle_oil_temp} the environmental and the nacelle versus time. For these two figures we note that the temperature exhibits seasonal behavior. Moreover, from the visual observation of the two plots in Figure~\ref{fig:env_nacelle_oil_temp}, we can conjecture that the environmental and the nacelle temperatures all behave very similarly. This was statistically confirmed by considering the correlation coefficient.

The vertical colored lines appearing in Figure~\ref{fig:env_nacelle_oil_temp} depict three major maintenance events, in red color:
\begin{itemize}
\item Large maintenance on the entire turbine, November 16, 2013;
\item A complete disassemble of the primary generator, March 28, 2014;
\item The deactivation of the primary generator, December 15, 2014;
\end{itemize}
and the considered period of in-control operational behavior, in green. The in-control period was considered to be before any major maintenance event and is statistically derived in Section \ref{sec:results}.

\begin{figure}
    \centering
    \begin{subfigure}[b]{0.45\textwidth}
        \includegraphics[width=\textwidth]{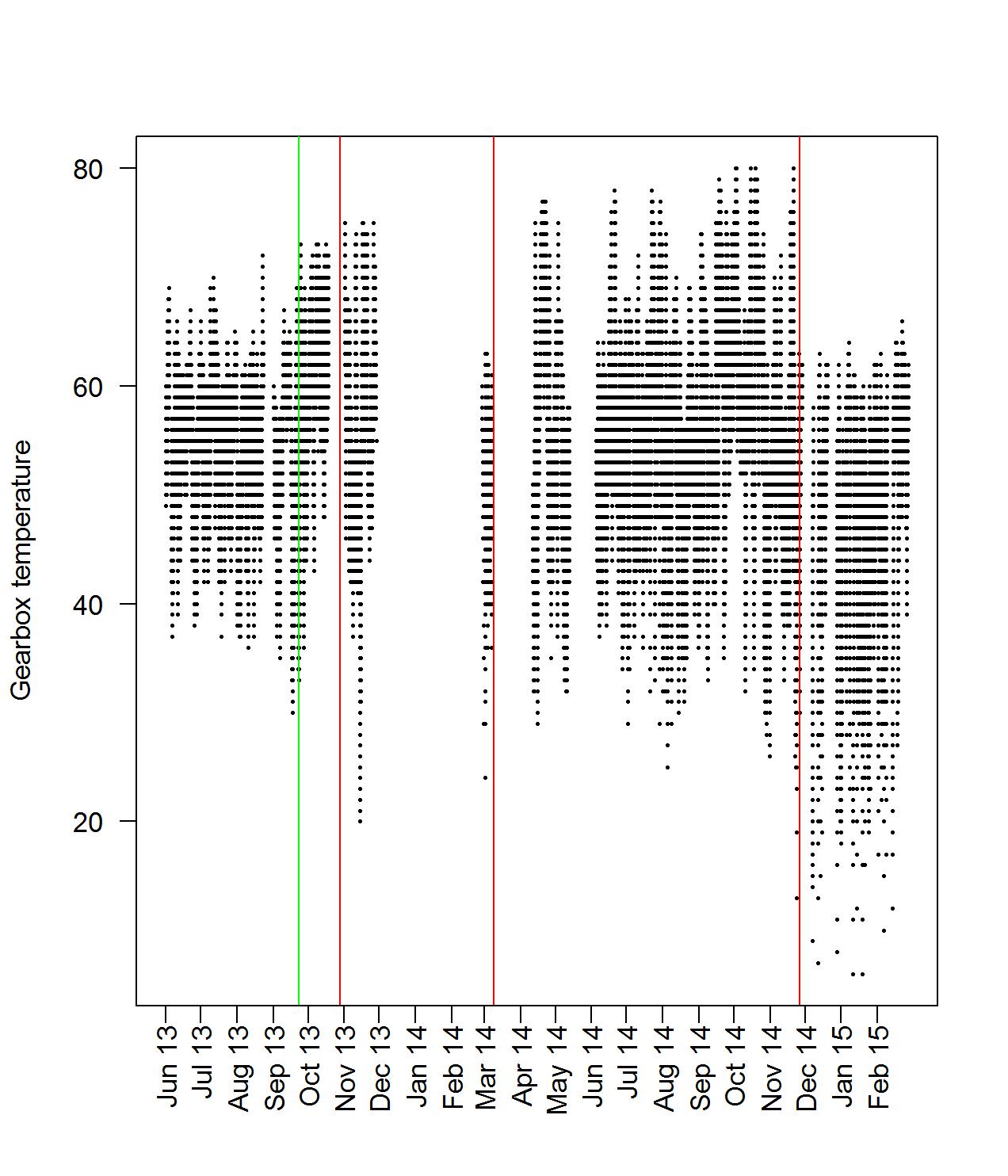}
        \caption{The gearbox  temperature}
        \label{fig:gbx_vs_time}
    \end{subfigure}
    ~
   \begin{subfigure}[b]{0.45\textwidth}
        \includegraphics[width=\textwidth]{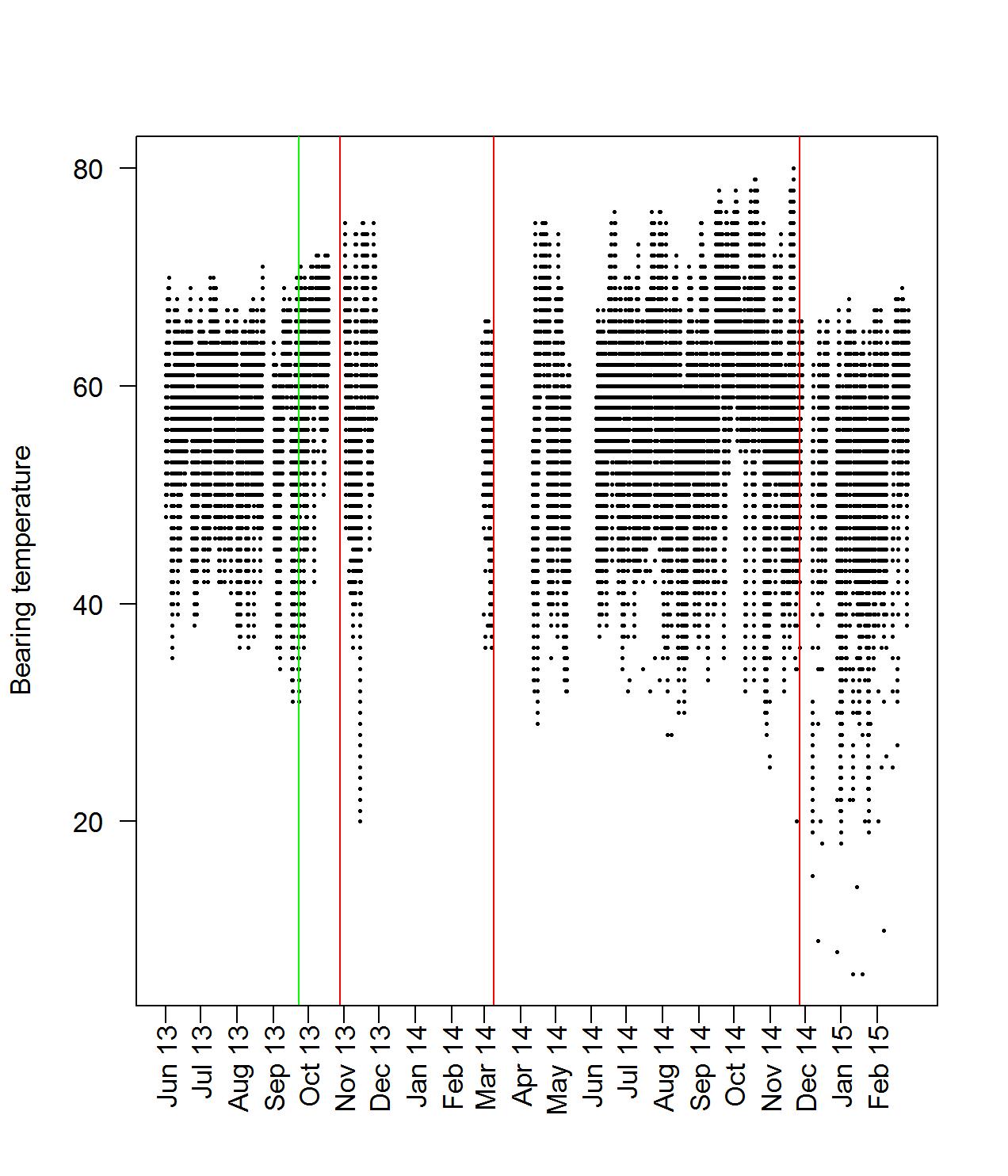}
        \caption{The  bearing temperature}
        \label{fig:brg_vs_time}
    \end{subfigure}
\caption{The gearbox  and the bearing temperature}
\label{plot:gbxbrg}
\end{figure}

\begin{figure}
    \centering
    \begin{subfigure}[b]{0.45\textwidth}
        \includegraphics[width=\textwidth]{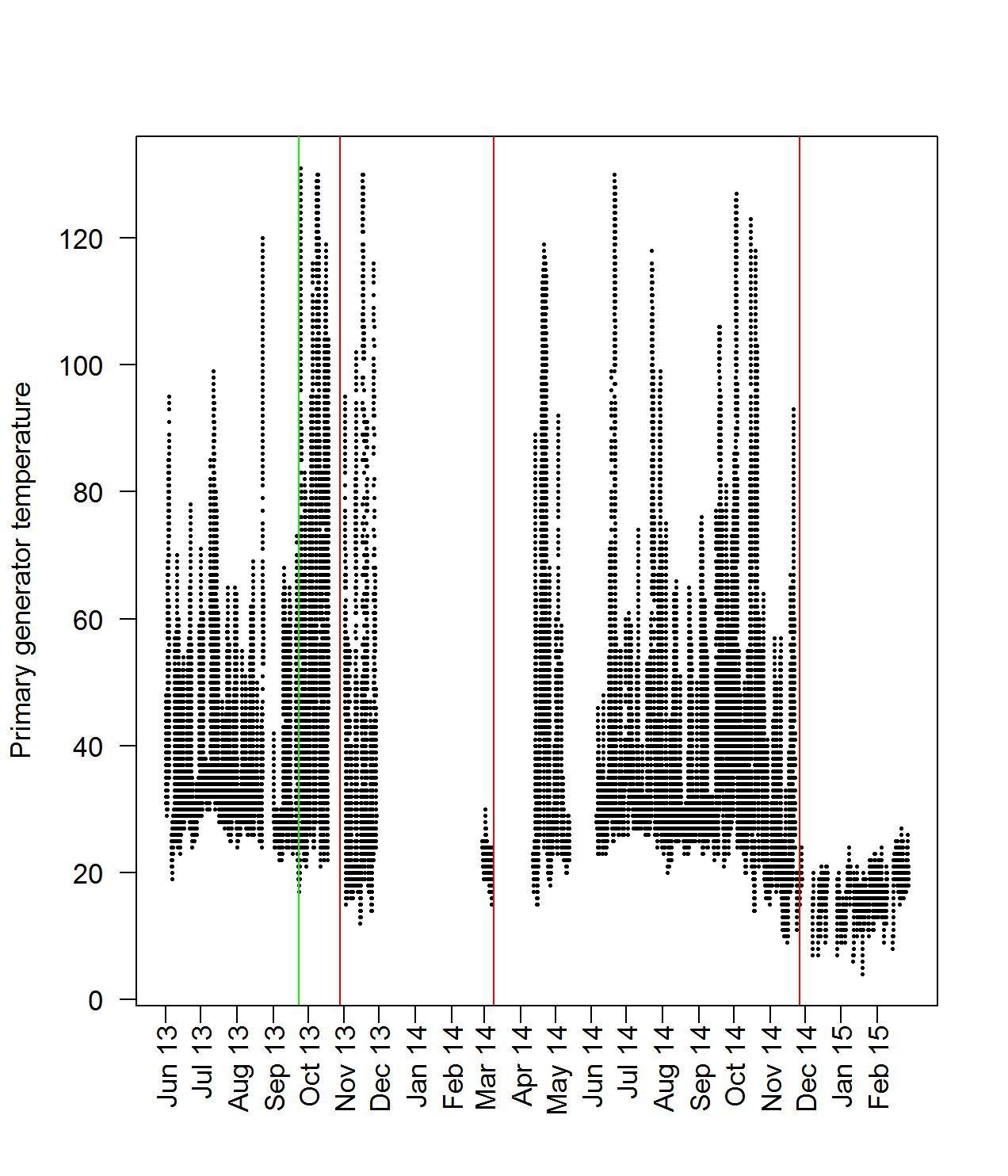}
        \caption{The primary generator  temperature}
        \label{fig:mgentemp_vs_time}
    \end{subfigure}
    ~
   \begin{subfigure}[b]{0.45\textwidth}
        \includegraphics[width=\textwidth]{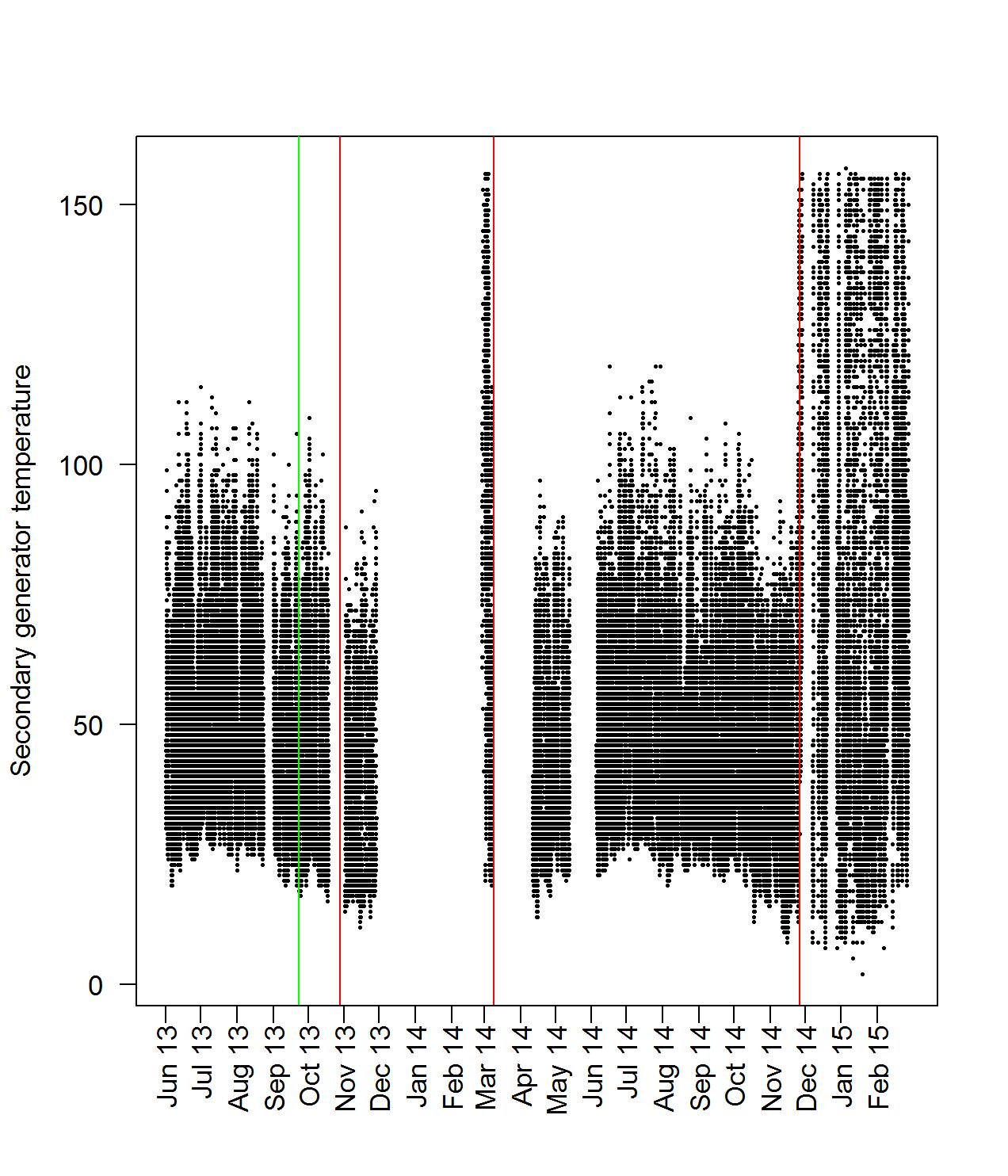}
        \caption{The  secondary generator temperature}
        \label{fig:mgentemp_vs_time}
    \end{subfigure}
\caption{The primary generator and the secondary generator temperature}
\label{plot:gen12}
\end{figure}

Next, in Figure~\ref{plot:gbxbrg}, the temperatures of the gearbox temperature and the bearing temperature are plotted. In Figure~\ref{plot:gen12}, the temperature of the
two generators are plotted against time.

The gearbox and bearing temperature behave very similarly. This is because of their physical connection that causes a direct heat transfer. If we look at the temperature of either generator, we observe that the minimum generator temperature seems to behave similarly to the environmental temperature, while the maximum temperature does not seem to be influenced by the environmental temperature. This can be explained by the fact that the generators only generate heat when they are in use. When they are not in use, the only source of heat comes from the nacelle cooling system, which is almost identical to the environmental temperature.

Other variables of interest include the power output, the wind speed, the rotor speed and the generator speed.
We see a sharp drop in power output around the end of 2014. This is explained by the failure of the primary generator. The maximum output of the secondary generator is lower than the output of the primary generator at higher wind speeds. We can visually identify low, moderate and high wind speeds through color coding, and plot the power output versus time.

The power seems to exhibit one pattern of behavior up to December 2014. We know from the maintenance logs that the primary generator was deactivated around December 2014, so this is why we see the sharp drop in power output. So it is imperative to distinguish, before any further analysis, which generator (secondary or primary) is in operation. Since there is no direct way for such an identification, we use the rotor speed. More concretely, we notice that the rotor speed has  a large concentration of values around 20 rotations per minute (RPM) and 25 RPM, and a big gap in between those two values. These large concentrations are created because each generator only works at one specific speed. The secondary generator generates power around 20 RPM and the primary generator generates power at 25 RPM.
Next, it is useful to look at some scatter plots, to  visually identify relationships between the various variables. To this purpose, we take a closer look at the relationship between the rotor speed, wind speed  and the power output.
\begin{figure}[htb]
\centering
\includegraphics[width=0.45\textwidth]{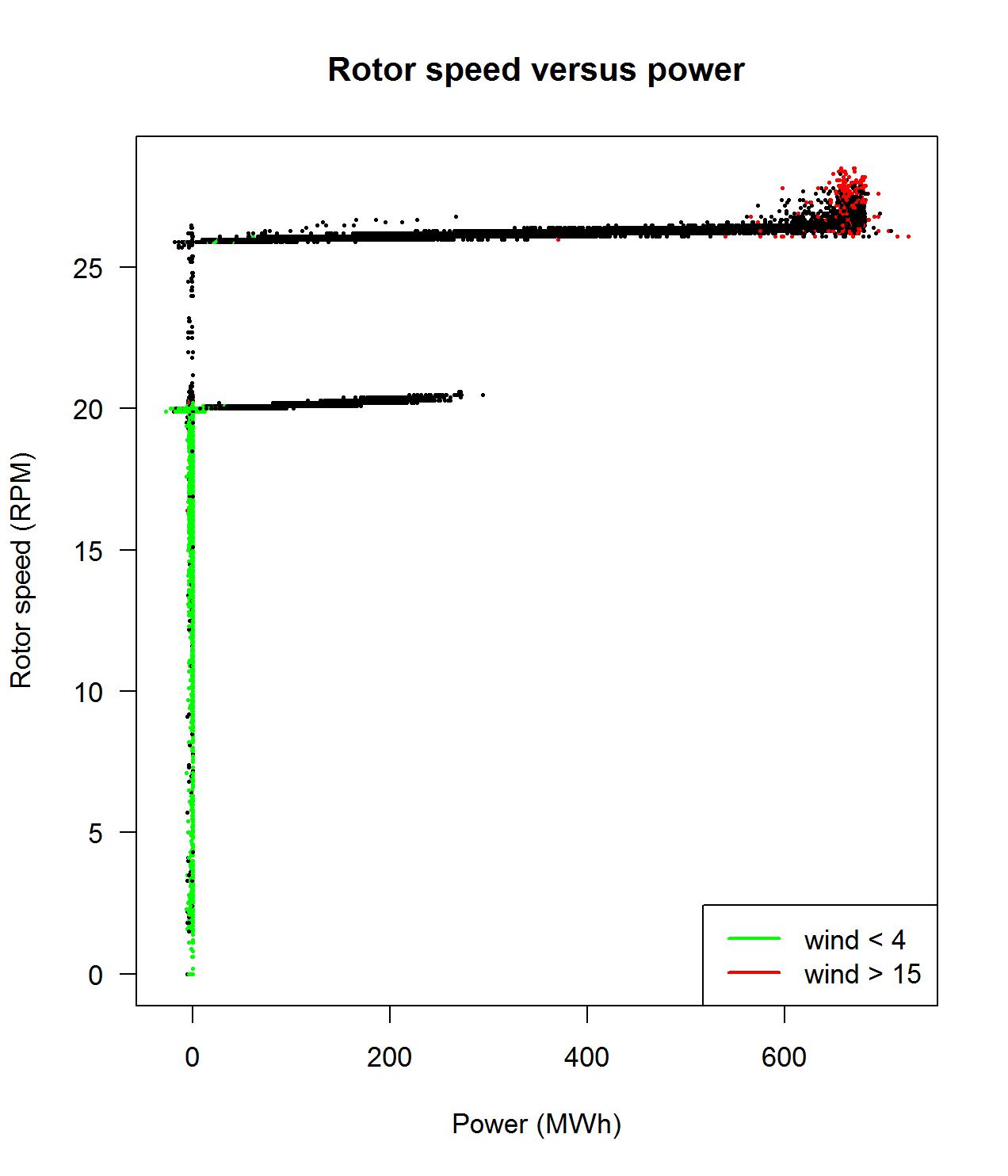}
\caption{Power output  versus rotor speed}
\label{img:powervrotorspeed}
\end{figure}

In Figure~\ref{img:powervrotorspeed}, the scatter plot of the rotor speed versus the power output is depicted. This plot should be read as follows: every paired observation of the rotor speed with the corresponding power output is depicted on the plot as a dot and is color-characterized based on the wind speed. In particular, if the wind speed is below 4 m/s, it is color-coded green in the plot. If the wind speed is between 4 m/s and 15 m/s it is color-coded black  and if the wind speed exceeds 15 m/s, it is color-coded red.
We see that all the points with a rotor speed of less than 19 RPM correspond very neatly to either a power output of 0 or a negative power output (consumption of power). This reinforces our hypothesis that a rotor speed below 19 RPM means that neither generator is in use. For the power output corresponding to either the primary generator or the secondary generator, the plots seem to indicate a cubic relation.
Note that the depicted plot resembles the letter ``F'', where the lower vertical line corresponds to power production  achieved by the secondary generator, while the upper vertical line corresponds to power production achieved by the primary generator. This is in accordance with the fact that the two generators work at two distinct speeds. This will later prove to be very useful, as there is no variable directly indicating which of the two generators is in use. Based on the rotor speed, we can determine which one of the two generators is in use. From this point onward we formulate 3 distinct cases and corresponding assumptions for the generator in use:
\begin{enumerate}
\item  When the rotor speed exceeds 25.8 rotations per minute (RPM), then we may assume that the primary generator is in use.
\item  When  the rotor speed exceeds 19 RPM but does not exceed 21 RPM, then we may assume that the secondary generator is in use.
\item  In all other cases there seems to be no power output, hence it seems unlikely that either generator is in use at those times.
\end{enumerate}

We draw the following conclusions from our exploratory data analysis:
\begin{itemize}
\item There is a very strong correlation between the environmental temperature and the nacelle temperature. To a lesser extend, this is also true for the oil temperature.
\item In order to model the critical components of the wind turbine, we combine together all relevant readings, e.g. temperature reading with vibration readings, after accounting for the variation coming from sources common to the operational process or the environment.
\item We can identify which generator is being used by looking at the rotor speed. A rotor speed around 20 RPM indicates that the secondary generator is in use, while a rotor speed of 25 RPM indicates that the primary generator is in use.
\item There seems to be a cubic relation between wind speed and power output. The theoretical curve for the Vestas V47 turbine is shown in Figure~\ref{plot:curve}.
\begin{figure}[htb]
\centering
\includegraphics[scale=1]{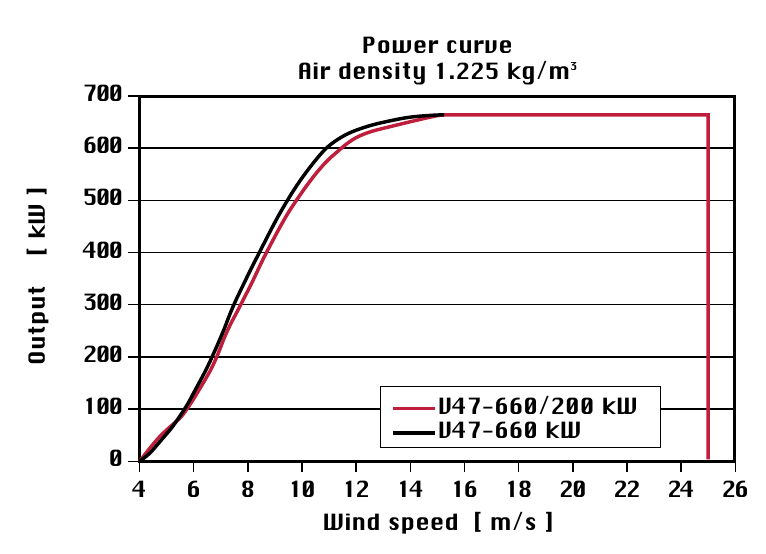}
\caption{Theoretical power curve for the Vestas V47 for air density 1.225 kg/m${}^3$, taken from \cite{Vestas-brochure}}
\label{plot:curve}
\end{figure}

\begin{figure}
    \centering
    \begin{subfigure}[b]{0.45\textwidth}
        \includegraphics[width=\textwidth]{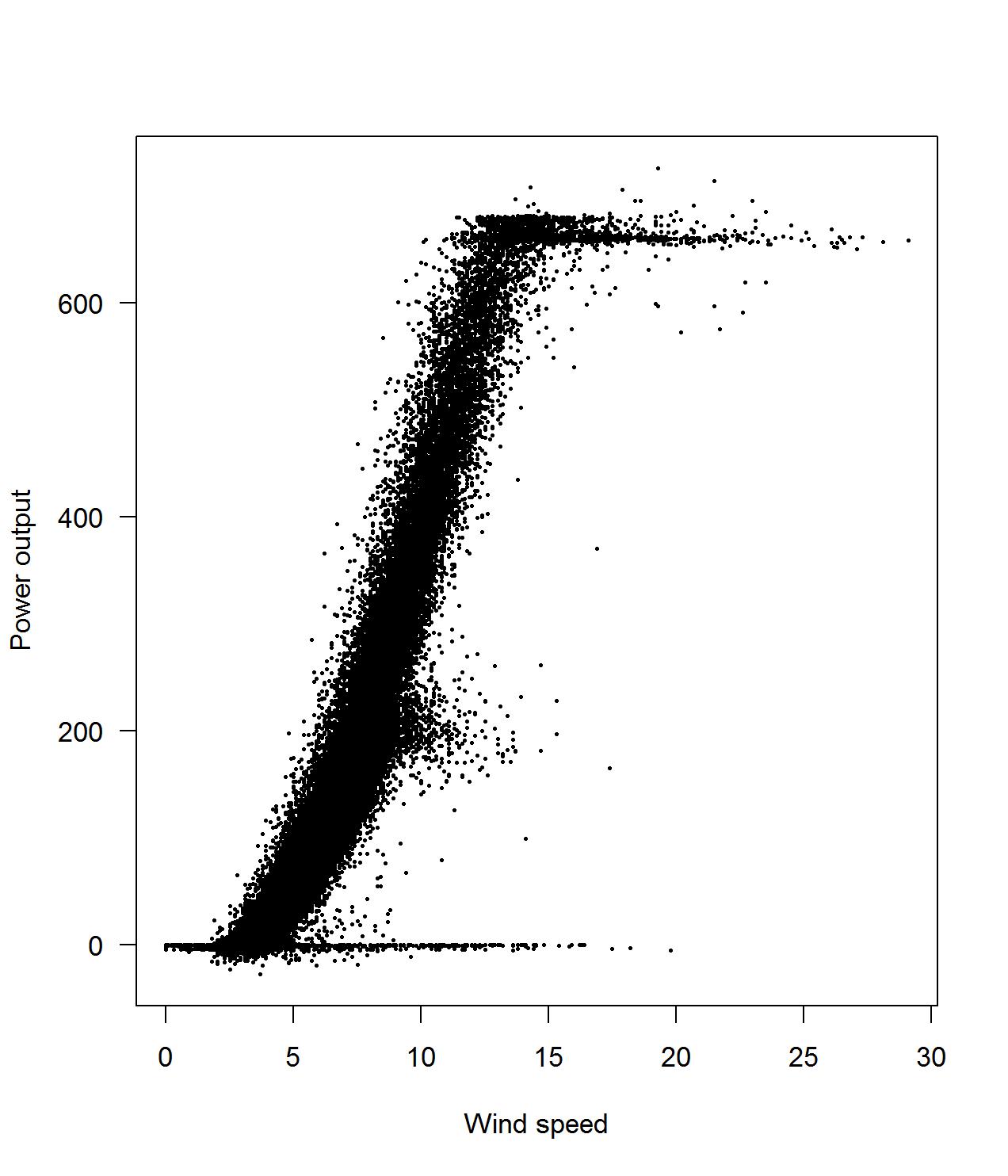}
        \caption{Power output}
        \label{fig:power_vs_time}
    \end{subfigure}
    ~
   \begin{subfigure}[b]{0.45\textwidth}
        \includegraphics[width=\textwidth]{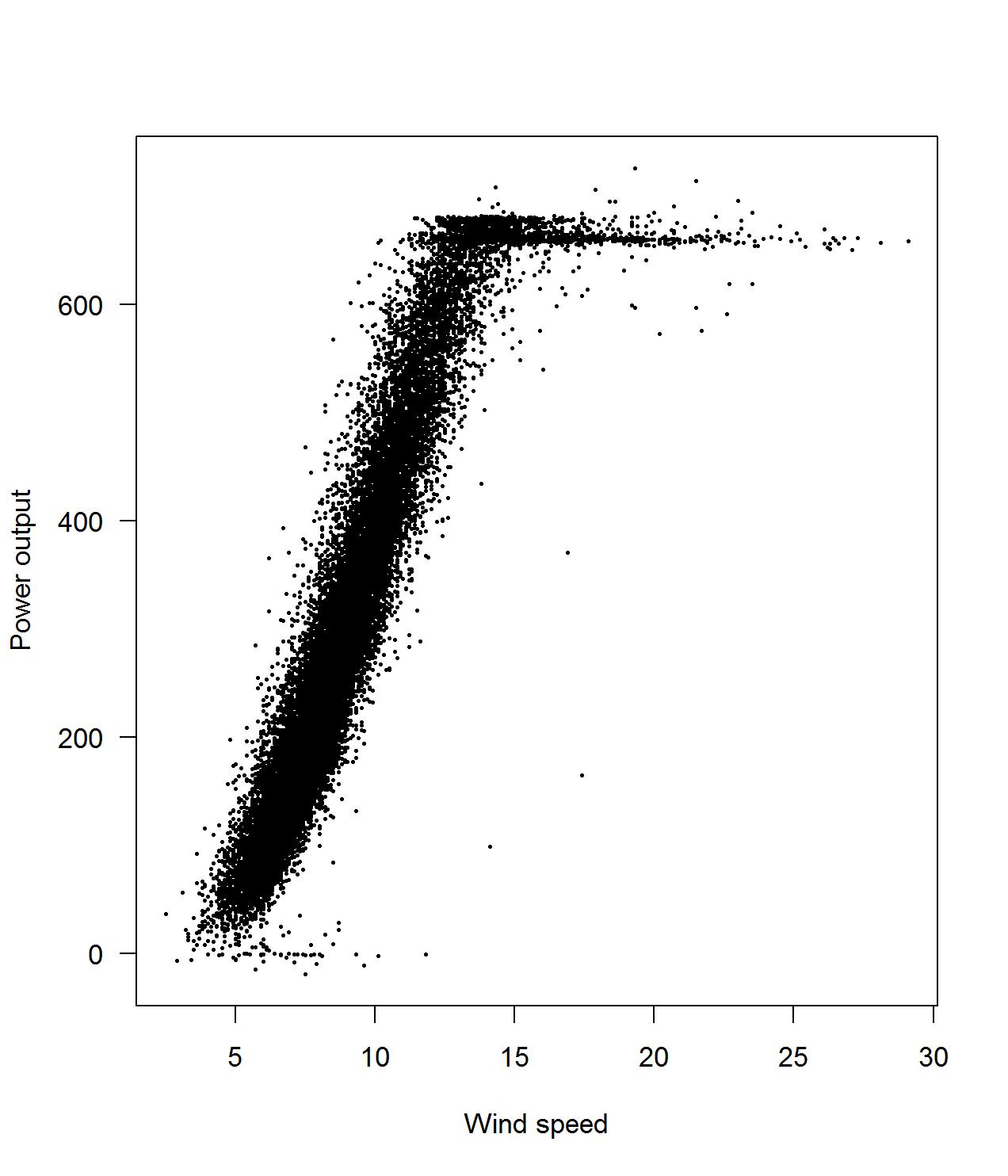}
        \caption{Power output for primary generator $(\text{RPM} \geq 25.8)$}
        \label{fig:powerm_vs_time}
    \end{subfigure}
    ~
   \begin{subfigure}[b]{\textwidth}
       \centering
        \includegraphics[width=0.45\textwidth]{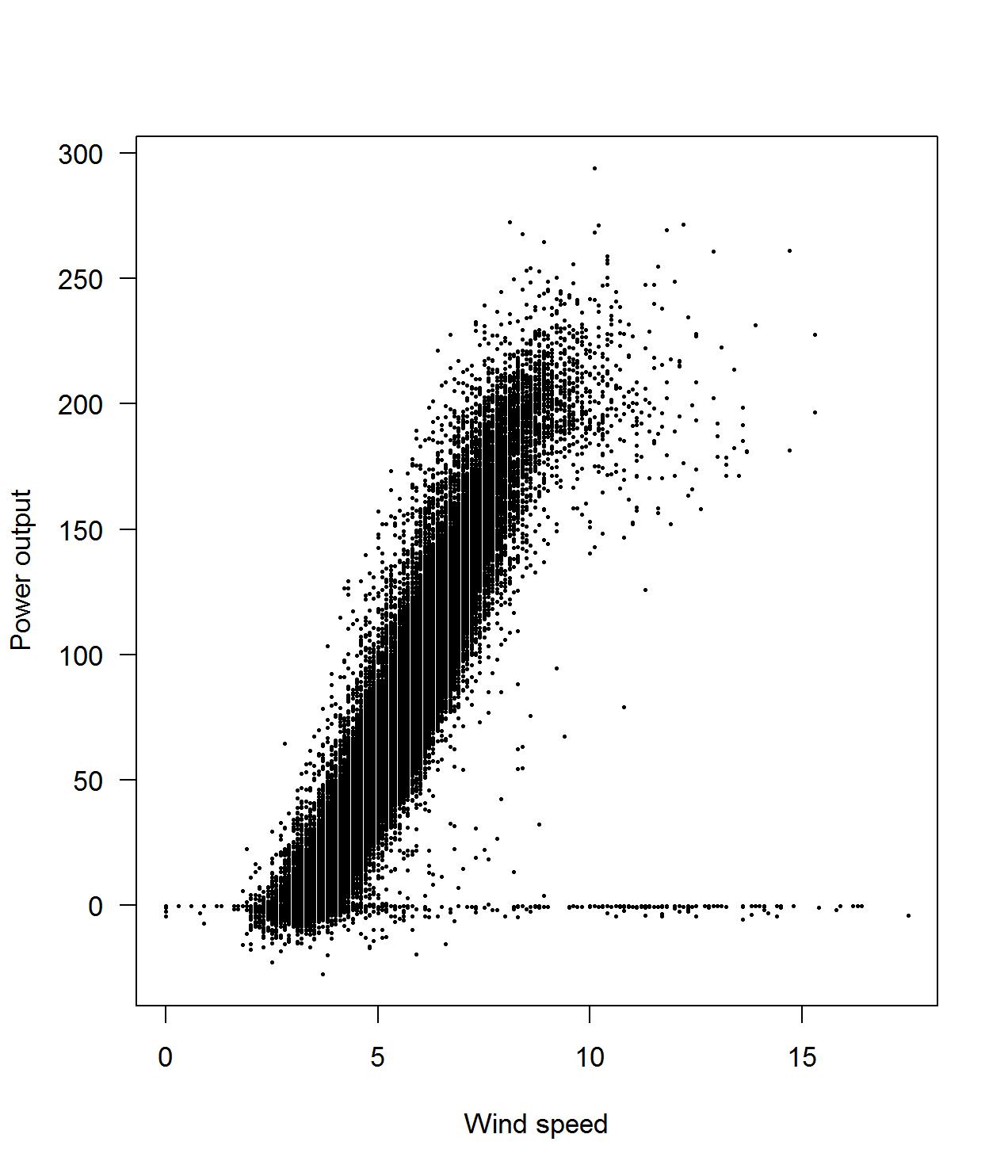}
        \caption{Power output for secondary generator $(19\leq \text{RPM}\leq 21)$}
        \label{fig:powers_vs_time}
    \end{subfigure}
\caption{Power output versus wind speed, split based on RPM corresponding to the primary generator $(\text{RPM} \geq 25.8)$ and the secondary generator $(19\leq \text{RPM}\leq 21)$}
\label{plot:powervwindspeed}
\end{figure}

Notice that the cubic shape of this curve resembles the scatter plot of the power output and wind speed in Figure \ref{plot:powervwindspeed}. In particular, note three important things about the curve
\begin{itemize}
\item Below $4$ m/s the turbine produces no power. There is insufficient torque exerted by the wind on the rotor to make it rotate and generate power. We say that $3.5$ m/s is the {\em cut-in speed}.
\item Between $4$ and 15 m/s, power behaves as a cubic function of wind speed. At 15 m/s, the limit of the generator power output is reached. After this point, the power output no longer increases. We say that 15 m/s is the {\em rated output speed}.
\item At wind speeds over 25 m/s, the forces exerted on the rotor are so great that there is a serious risk of structural damage. Therefore, the breaking system brings the system to a halt. We say that 25 m/s is the {\em cut-out speed}.
\end{itemize}

Between the cut-in speed and the rated output speed, the estimated output power can be shown to be estimated by the following formula
\begin{equation}
P_w = \frac{1}{2} \rho \, A_r \, c_p \, u_w^3,
\end{equation}
where $P_w$ is the power in watts, $u_w$ is the wind speed at hub height upstream the rotor in meters per second, $\rho$ is the aerial density in kilograms per cubic meter,    $c_p$ is the performance coefficient or power coefficient, and $A_r$ is the area covered by the rotor in square meters, see \cite[Equation~(3)]{Slootweg}. This is in accordance with Figure \ref{plot:powervwindspeed}.
\end{itemize}

\section{Approach}\label{sec:approach}
In order to arrive at adaptive thresholds that give timely warnings while keeping false warnings at a minimum, we developed an approach consisting of the following steps:
\begin{description}
\item[Step 1.] Determine a baseline period with normal operational  behavior, aka the in-control period;
\item[Step 2.] For the period of Step 1 create a linear regression model for the parameter of interest;
\item[Step 3.] Based on the regression model of the previous step, determine adaptive thresholds for adjusted parameters by considering the residuals of the regression model;
\item[Step 4.] Monitor the residuals of the regression models separately using the adaptive thresholds of Step 3.
\end{description}

From the problem statement, recall that the purpose of the case study is to produce models which will timely predict  failures of any of the wind turbine parts or the turbine as a whole. The way to do this, is to detect deviations from what we consider to be ``normal behavior'' in these models. To identify what qualifies as normal behavior and to determine what deviates from normal behavior, we will employ techniques from a group of methods known as statistical process control (SPC).

SPC is a method of quality control which uses statistical methods. It is applied in order to monitor and control a process by looking at deviations from normal behavior. SPC originated in the manufacturing industry, as a way to monitor the quality and consistency of the manufactured items. It goes back to Walter Shewhart, who introduced the concept in 1924.

The standard setting for control charts is to collect small groups of observations at distinct time points. These groups usually occur in a natural manner. They could consist, for example, of items that were manufactured simultaneously. These groups are called rational subgroups. Typically, the mean and/or the standard deviation of each group is assessed for ``normal'' behavior. In SPC-terminology, normal behavior is called in-control, while deviations from normal behavior are called out-of-control. Control is actually a misnomer in this case. It is not about control in the engineering sense of the word, like in feedback control. The processes are actually about monitoring. The word ``control'' is still used for historical reasons.

The main idea behind SPC is that there is variation in every process, but that this variation can be described as being one of two types: in every process, there is intrinsic variation. This is variation that always occurs due to the nature of the process, which will never be 100 \% deterministic. Shewhart called this variation due to common causes. This type of variation always occurs, even in an in-control situation. Secondly, there is variation due to special causes. This is the type of variation that defines an out-of-control situation, and the purpose of SPC is exactly to detect variation of this type. The reason rational subgroups are used is to get an unbiased estimator of the process variance. This is then compared to the level of variation which one expects in an in-control situation (the variation due to common causes). The simplest way to make these ideas operational is through the so-called Shewhart control chart which is a time plot of individual observations or (typically) the mean of a group of observations. Such a chart is a way to visualize the variation of a process and to compare it to what one expects due to common causes.
The control limits are usually set at $\bar{x} \pm 3 \frac{\hat{\sigma}}{\sqrt{n}}$, where $\bar{x}$ and $\hat{\sigma}$ are determined from historical data. This data has come from a period, that is known to be in-control, otherwise known as the start-up period. This period is usually called phase I. Other choices exist, such as the range estimator for the standard deviation. This is turned into an unbiased estimator by using a multiplicative constant. The actual monitoring phase is called phase II (also called on-line phase). The control chart gives a signal when the mean of a rational subgroup is outside the control limit.

Not everything defined in the standard setting above is applicable to our situation. In our approach we do not  monitor the observed parameters like temperatures and vibrations but monitor the residuals from the corresponding regression models, i.e. we monitor values adjusted for external influences.  we develop regression models and monitor the (non-standardized) residuals. The idea to use regression models in control charting is usually attributed to \cite{mandel1969regression} for the simple linear regression case.  It has appeared in several settings like  tool wear charts,   in multivariate SPC ( \cite{hawkins1991multivariate,hawkins1993regression} where each response is regressed on the other responses) or in cause-selecting charts in multistage processes , software maintenance: \cite{haworth1996regression} and structural health monitoring \cite{ozkul2011regression}. The theoretical properties have received relatively little attention in the literature; for the important issue of the effect of parameter estimation in control charts based on simple linear regression we refer to \cite{shu2005effects}. Most papers use ordinary linear regression, but two alternatives have been studied: support vector machines (\cite{gani2010svmregressioncontrolcharts} and L1-regression (LAV) \cite{ozkul2011regression}. The idea to monitor residuals of  regression models should not  be confused with profile monitoring, where the observations are profiled modeled as regression models themselves, see e.g. \cite{noorossana2012statistical}.

Another difference with the standard setting is that there are no natural subgroups because we are observing a time series. This is called control chart for individuals in the SPC literature, see \cite{MontgomerySPCbook}. In this case one cannot use the sample variance of the rational subgroup as it is undefined for groups of size 1. As a proxy, the moving range $\text{MR}_i=|x_i+1 - x_i|$ is used as an estimator for the short term standard deviation. This is important in case of a gradual shift in the mean.

\section{Results}\label{sec:results}
\begin{figure}[htb]
\centering
\includegraphics[width=0.45\textwidth]{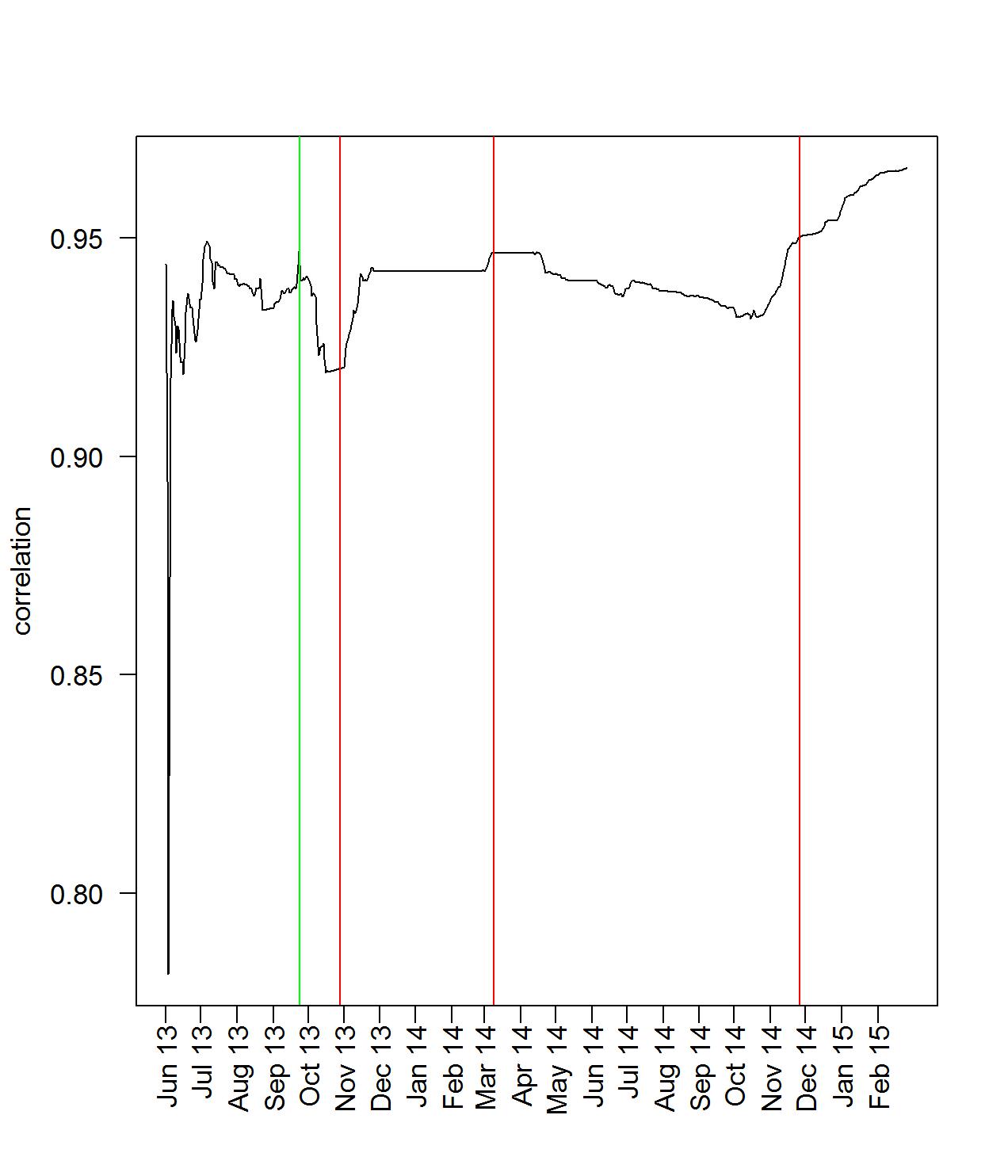}
\caption{Correlation coefficient of nacelle temperature and environmental temperature as function of time}
\label{fig:correlation_time}
\end{figure}

We now apply the above methods to several variables, namely:
\begin{itemize}
\item The nacelle temperature;
\item The primary generator temperature;
\item The primary generator vibration readings.
\end{itemize}
For each of these variables, we create a regression model and construct a method to check for warnings.
In order to identify the behavior of the aforementioned readings as out-of-control, we need first to identify an in-control period. This is chosen as a period without any major mechanical errors, at least based on the maintenance and event logs. We know that a major maintenance event took place on November 16, 2013, so we can already take this date as an upper bound of the in control period. However, out of control behavior most likely started earlier than that date. We would like to identify a period that is as long as possible, and without any negative influence from the events that lead to the maintenance activity on November 16, 2013. In order to achieve this, we look at the correlation coefficient of the environmental temperature and the nacelle temperature. More concretely, we consider the correlation coefficient of these two variables up to a certain point in time, say $t$, and we vary the value of $t$. Our objective is to determine the $t$ that maximizes the correlation coefficient. Let
\begin{equation*}
\rho (t) = \rho_{n,e}(T_0, t), T_0 \leq t \leq T_M,
\end{equation*}
where $\rho_{n,e}(T_0, t)$ is defined as the correlation coefficient of the environmental temperature and the nacelle temperature from time $T_0$ up to time $t$, see  Figure~\ref{fig:correlation_time}.
As  expected, the correlation values as a function of $t$ initially fluctuate, then stabilize, and show a slight decrease leading up to the maintenance event on November 16, 2013. The maximum is obtained on October 12, 2013. Thus, from this point onward we consider the period from  $19/06/2013$ up to $12/10/2013$ as our in-control period.

\begin{figure}
    \centering
    \begin{subfigure}[b]{0.45\textwidth}
        \includegraphics[width=\textwidth]{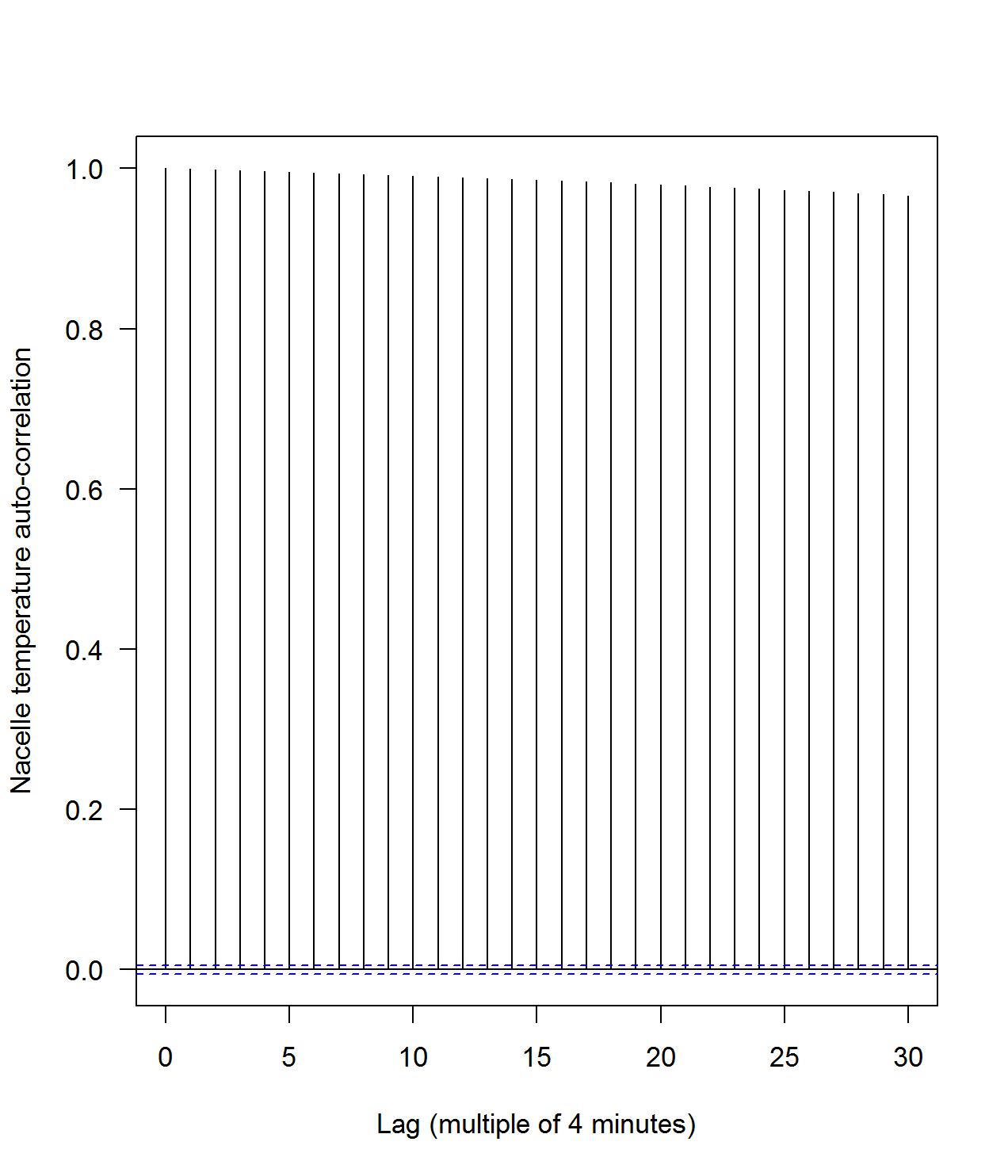}
        \caption{Autocorrelation for the nacelle temperature with 4 minute lag}
        \label{fig:nac_aut_4m}
    \end{subfigure}
    ~
   \begin{subfigure}[b]{0.45\textwidth}
        \includegraphics[width=\textwidth]{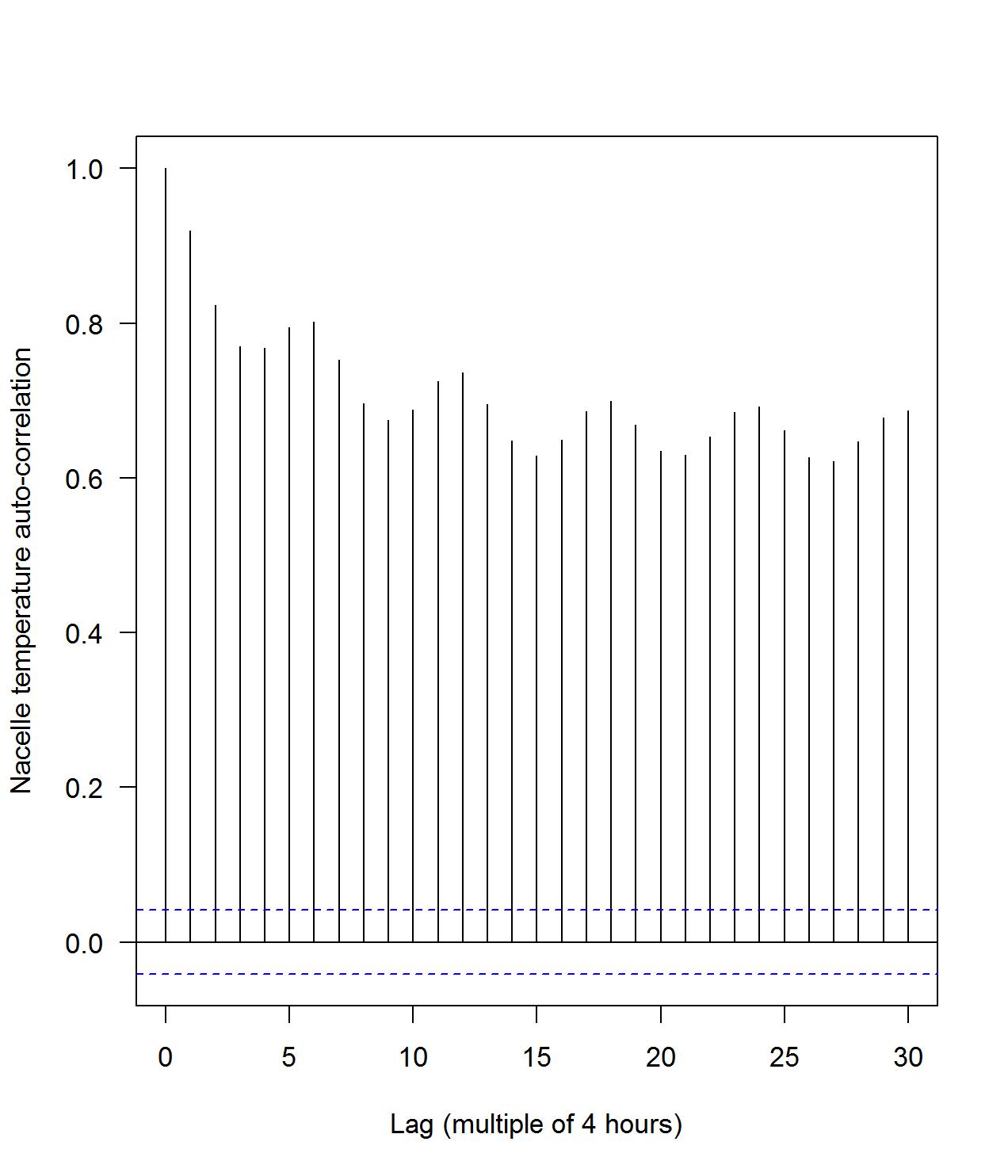}
        \caption{Autocorrelation for the nacelle temperature with 4 hour lag}
        \label{fig:nac_aut_4h}
    \end{subfigure}
\caption{Auto-correlation function for the nacelle temperature using the entire data set in the left plot, and the reduced (4h-intervals) in the right plot} \label{fig:autocor}
\end{figure}

The various readings are collected with a frequency of four minutes. While this provides us with a great amount of information, it also leads to issues due to the inherent autocorrelation of the measurements. In order to reduce this, we subsample from the original data set with a frequency of 4 hours. This should reduce the autocorrelation. In fact, this is evident from the autocorrelation plots for the nacelle temperature, see Figure~\ref{fig:autocor}.

\subsection{Model for the nacelle temperature}
We first try our approach on the nacelle temperature. Since the nacelle contains all the other components, its temperature should reflect a change in behavior if any of the components contained in the nacelle exhibits a high temperature. Furthermore, it has the added benefit of being the easiest to model, since, as we've seen, it behaves similarly as the environmental temperature. In fact, in building a regression model for the nacelle temperature it turns out that the environmental temperature is the only explanatory variable needed. First, we consider a baseline period, aka the in-control period, during which we construct the regression model. In Section \ref{sec:results} we determined the in-control period to be the period till 12/10/2013. This is the period for which we construct our regression model for the nacelle temperature. Thereafter, we use this regression model for predicting the nacelle temperature for the entire period. If the actual nacelle temperature is statistically different from the predicted one, this is an indication for a warning.

\begin{figure}
    \centering
    \begin{subfigure}[b]{0.45\textwidth}
        \includegraphics[width=\textwidth]{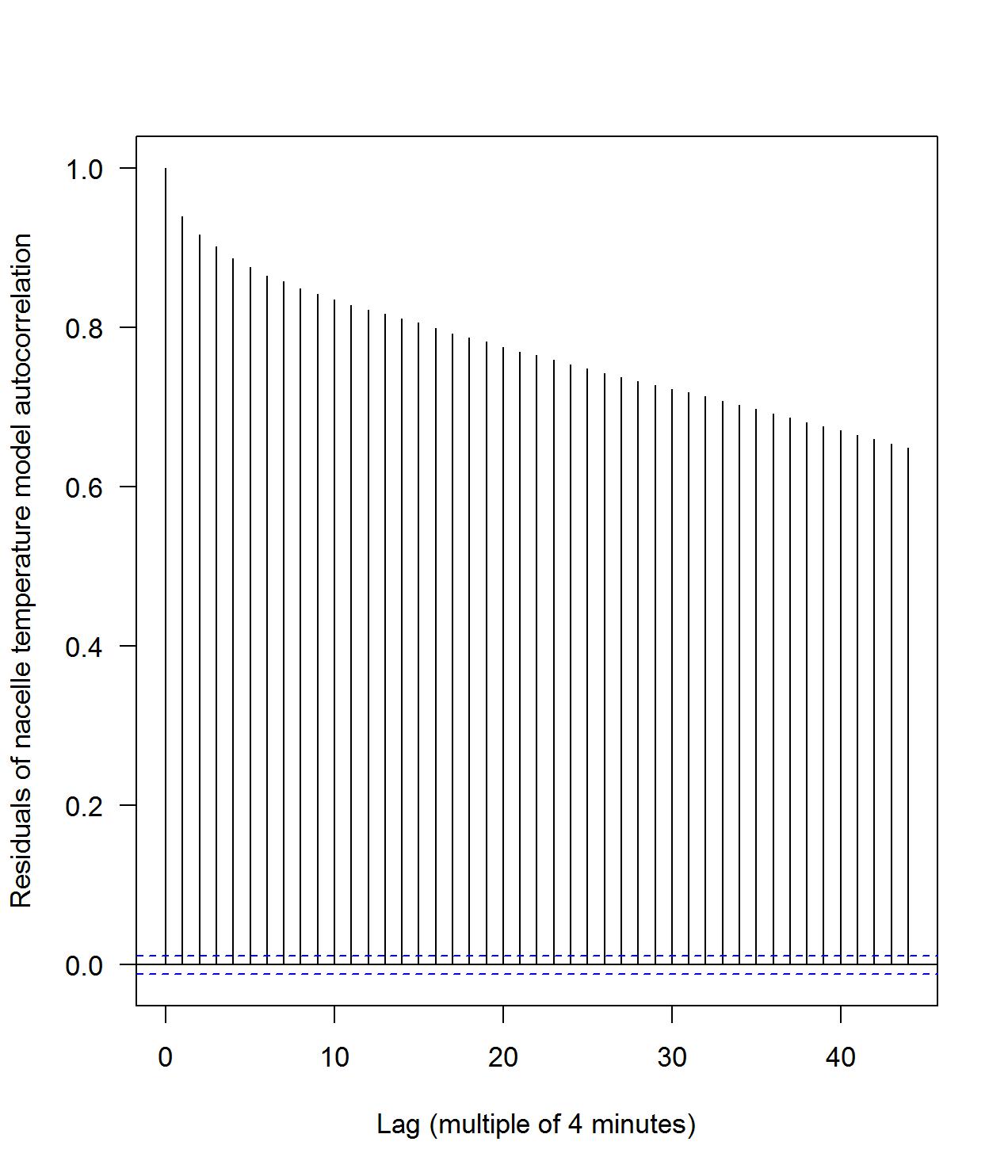}
        \caption{Autocorrelation for the residuals of the nacelle temperature model with 4 minute lag}
        \label{fig:rnac_aut_4m}
    \end{subfigure}
    ~
   \begin{subfigure}[b]{0.45\textwidth}
        \includegraphics[width=\textwidth]{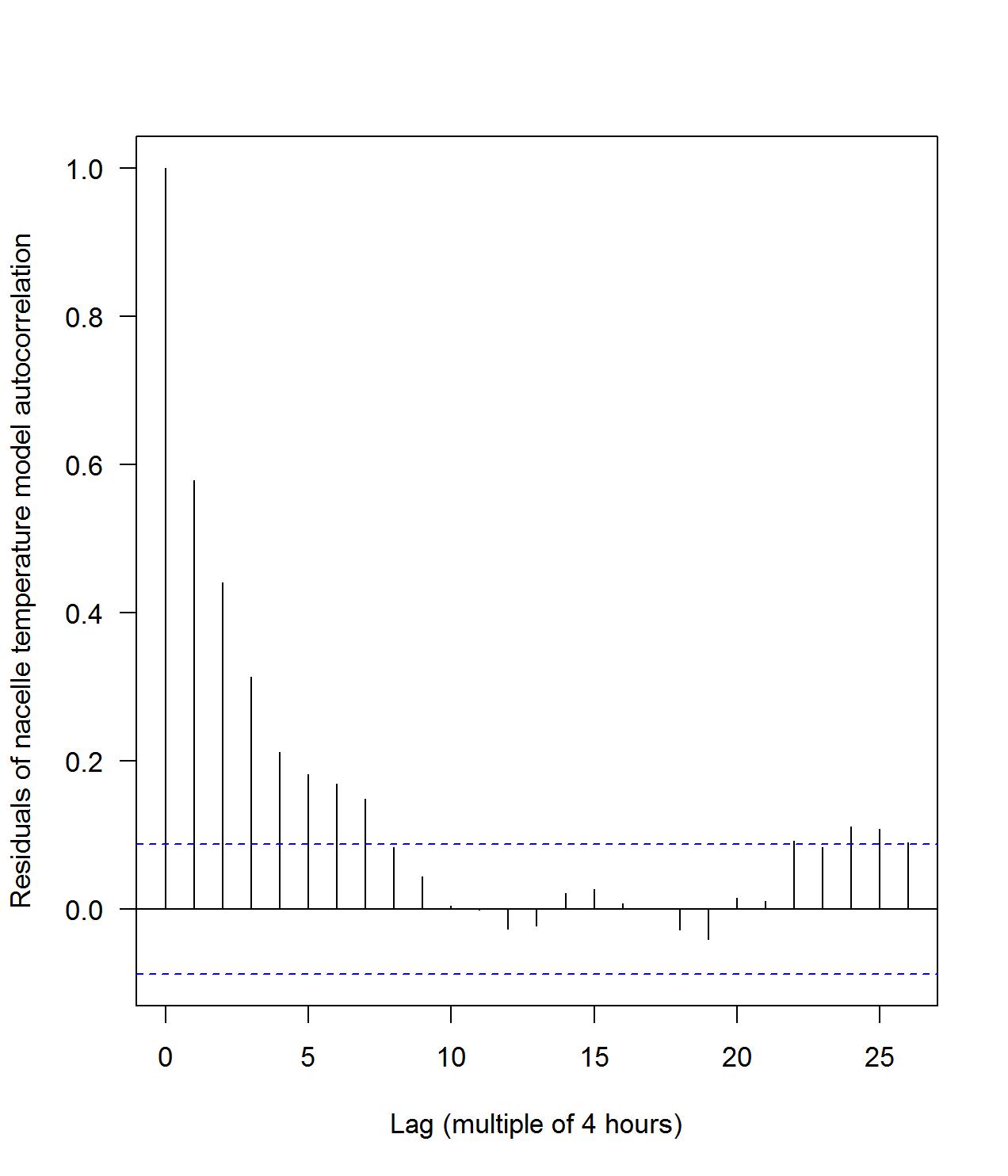}
        \caption{Autocorrelation for the residuals of the nacelle temperature model with 4 hour lag}
        \label{fig:rnac_aut_4h}
    \end{subfigure}
\caption{Autocorrelation of the residuals of the nacelle temperature using the entire data set in the left plot, and the reduced (4h-intervals) in the right plot}
\label{fig:autocor_NC}
\end{figure}

Next, we compare whether a regression model based on data every 4 hours or every 4 minutes would give us the best results. To this end, we create two preliminary regression models. One based on an in-control period and using all data points within that period, and one using a subset of the data within that period, with measurements  every 4 hours instead of every 4 minutes. Note that the autocorrelation plot of the residuals of these two regression models, depicted in Figure~\ref{fig:autocor_NC}, indicate that the choice of the sub-sampling every 4 hours is more consistent with the requirements of regression analysis and suffers less from the inherent autocorrelation. and visually inspect it to see which model performs best in terms of autocorrelation.

Since there is only one explanatory variable in the model, in this case it is not necessary to use stepwise- or all subset regression to improve the selection of the explanatory variables. The coefficients of the nacelle temperature model are listed in Table ~\ref{tab:nacellemodel_4h}.

\begin{table}[hbt]
\centering
\begin{tabular}{|r | r|r|r|r|}
  \hline
\multicolumn{1}{|r}{} & \multicolumn{1}{r|}{Estimate} & \multicolumn{1}{r|}{Std. Error} & \multicolumn{1}{r|}{t-value} & \multicolumn{1}{r|}{p-value} \\
\hline
(Intercept) & 7.54899  &  0.57528 &  13.12 &  $<$2e-16 \\
   \hline
EnvTemp &  0.94560   & 0.02534 &  37.31  & $<$2e-16 \\
   \hline
\end{tabular}
\caption{Regression model of the nacelle temperature, 4 hour intervals}
\label{tab:nacellemodel_4h}
\end{table}

\begin{figure}
\centering
\includegraphics[width=0.45\textwidth]{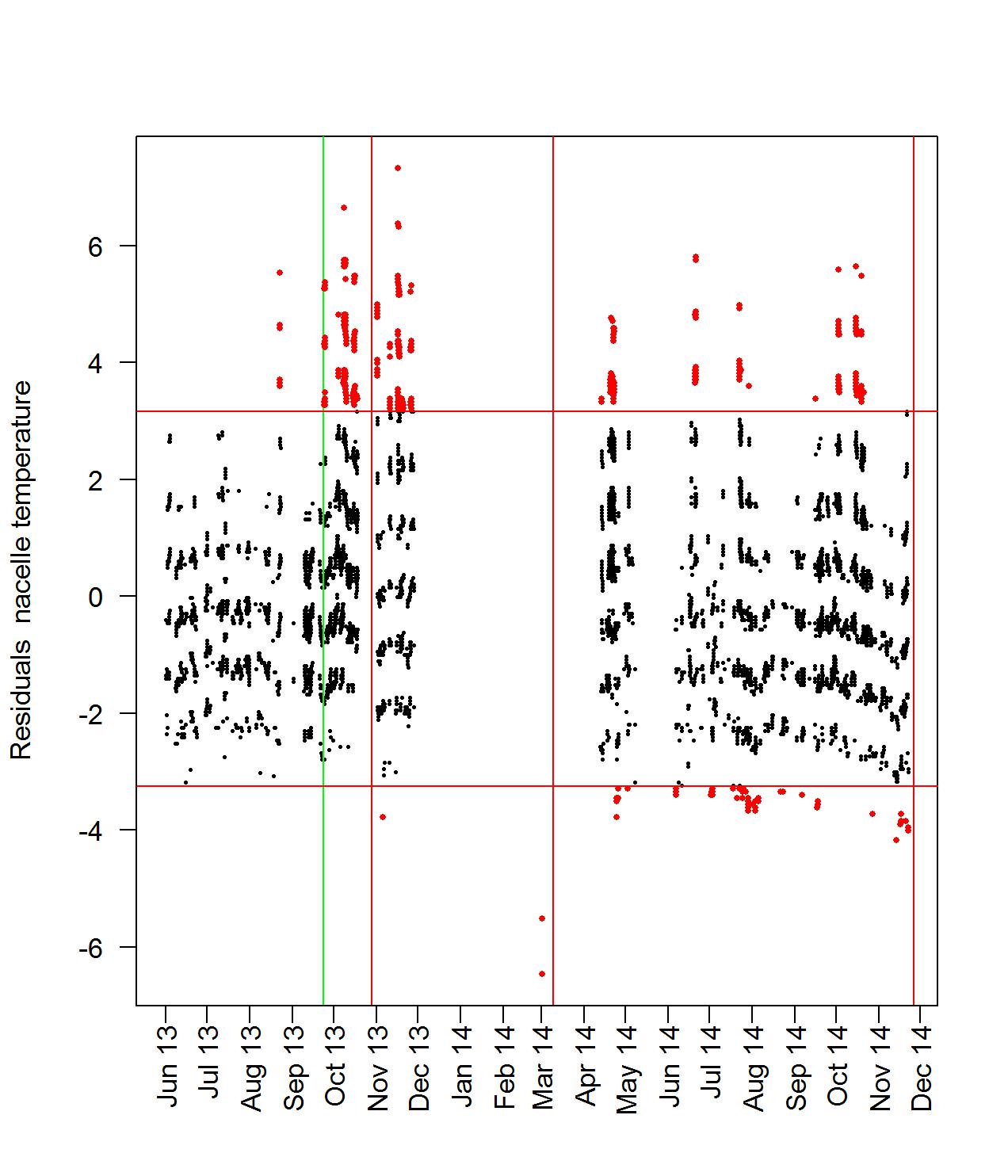}
\caption{Shewhart control chart for the nacelle temperature when using an in-control period to calculate the regression model, and using 4 minute intervals}
\label{shewhart:nacelle}
\end{figure}

Using this regression model, which is based on the in-control period, we can predict point estimates for the nacelle temperature for the entire data period and also calculate the difference between the actual values and the predicted ones. This difference is referred from now on as residuals, and these residuals are used to calculate the control limits for a Shewhart chart, which allows us to see which data points behave out-of-control. The control limits are calculated using data from the in-control period. Any data acquired after the control limit is not included in the computation of the control limits. In particular the control limits are -3.256612 and 3.159216. Any residual outside these limits is characterized as out-of-control. The Shewhart chart is shown in Figure~\ref{shewhart:nacelle}. In total, 2158 measurements are outside the control limits, out of a total of 30775, or approximately 7,01\%. This is much more than is to be expected from a well-behaved system using a 99,75\% confidence level, and this is definitely an indication that things are not functioning as expected.

\begin{figure}
\centering
\includegraphics[width=0.45\textwidth]{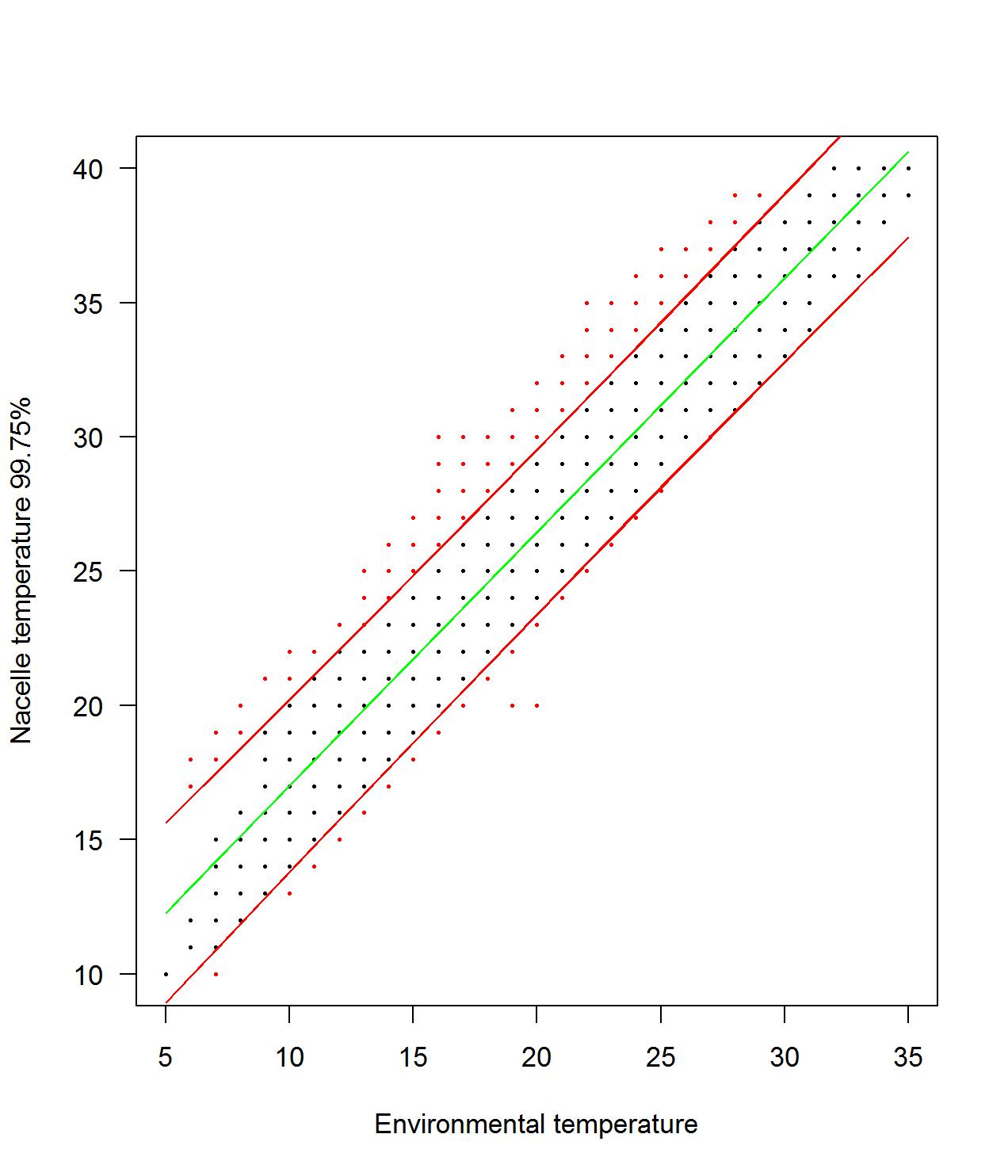}
\caption{Confidence interval for the nacelle temperature when using an in-control period to calculate the regression model, and using 4 minute intervals}
\label{nacelleplot_incontrol}
\end{figure}

For understanding purposes of how the Shewhart chart works, we create a scatter plot of the nacelle temperature versus the environmental temperature. In this scatter plot, we depict all measurements that are characterized as out-of control with red, see Figure~\ref{nacelleplot_incontrol}. Furthermore, we depict the upper and lower control limits as red lines on the scatter plot. Note that  all out-of control points lie outside the control limits lines. Since a defect is more likely to cause an increase in temperature than a decrease, this might indeed be the effect of a malfunctioning of some part of the turbine. We also see that most out-of-control measurements occur at moderate temperatures rather than at temperatures which are extremely high or extremely low. The lack of out-of-control measurements at lower temperatures may however also be caused by the period that we identified as in-control. This period, which runs from June to October, did not have many low temperatures, so the predictions may be less accurate in months during which the environmental temperature is low. An in-control period that covers both warm and cold months might yield even better results, but unfortunately this is not possible with the data that we currently have.

\begin{figure}
\centering
\includegraphics[width=0.45\textwidth]{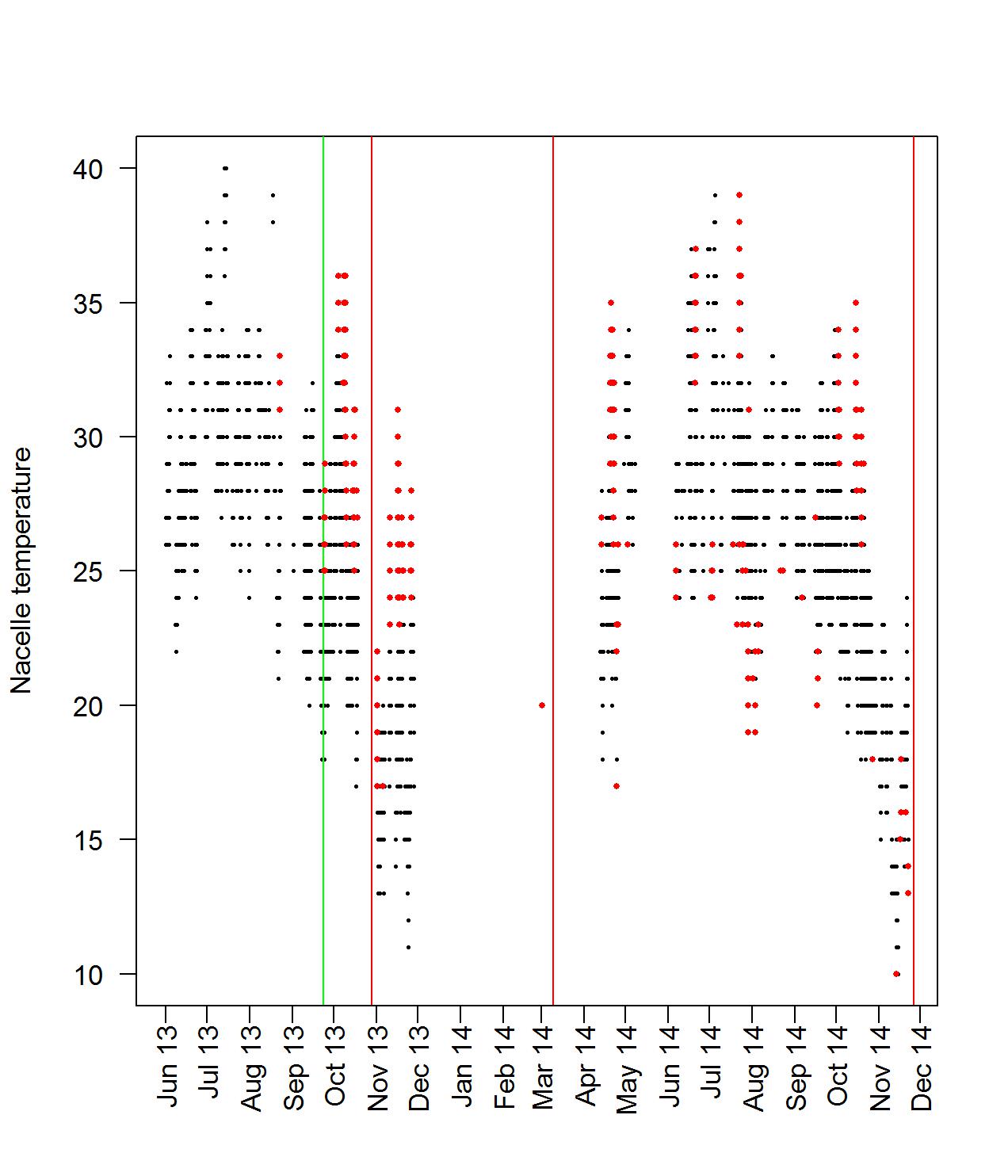}
\caption{Nacelle temperature versus time. The points which are identified as warnings are color-characterized in red.}
\label{plot:nacelletime}
\end{figure}

For illustration purposes we color-characterize all warnings (out-of-control measurements) derived from the Shewhart chart on a plot of the nacelle temperature versus time, see Figure \ref{plot:nacelletime}. We expect in such a figure to see a concentration of out-of-control measurements before maintenance activities. This is clearer in the next two models.


\subsection{Model for the generator temperature}
\begin{figure}
    \centering
    \begin{subfigure}[b]{0.45\textwidth}
        \includegraphics[width=\textwidth]{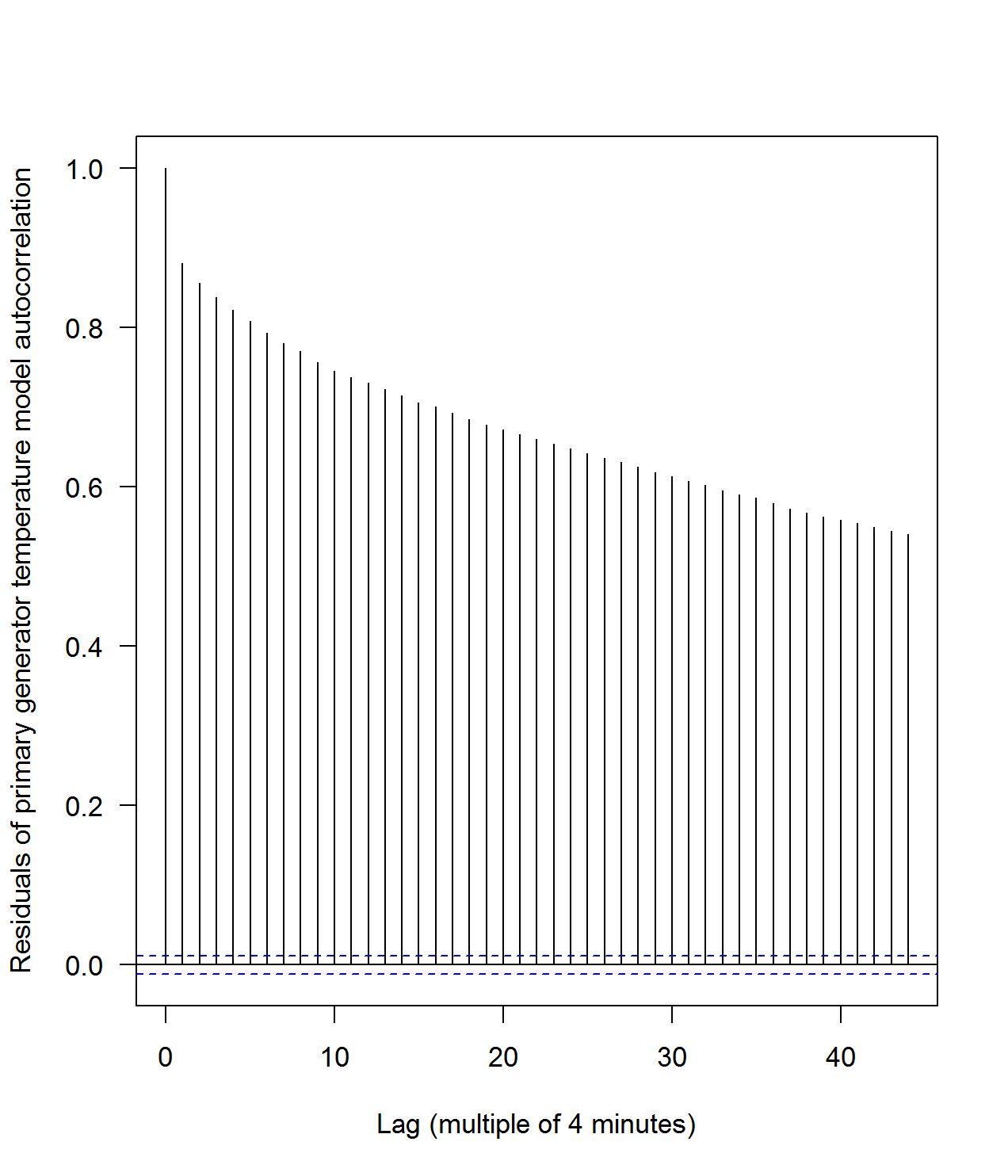}
        \caption{Autocorrelation for the residuals of the primary generator temperature model with 4 minute lag}
        \label{fig:rgentemp_aut_4m}
    \end{subfigure}
    ~
   \begin{subfigure}[b]{0.45\textwidth}
        \includegraphics[width=\textwidth]{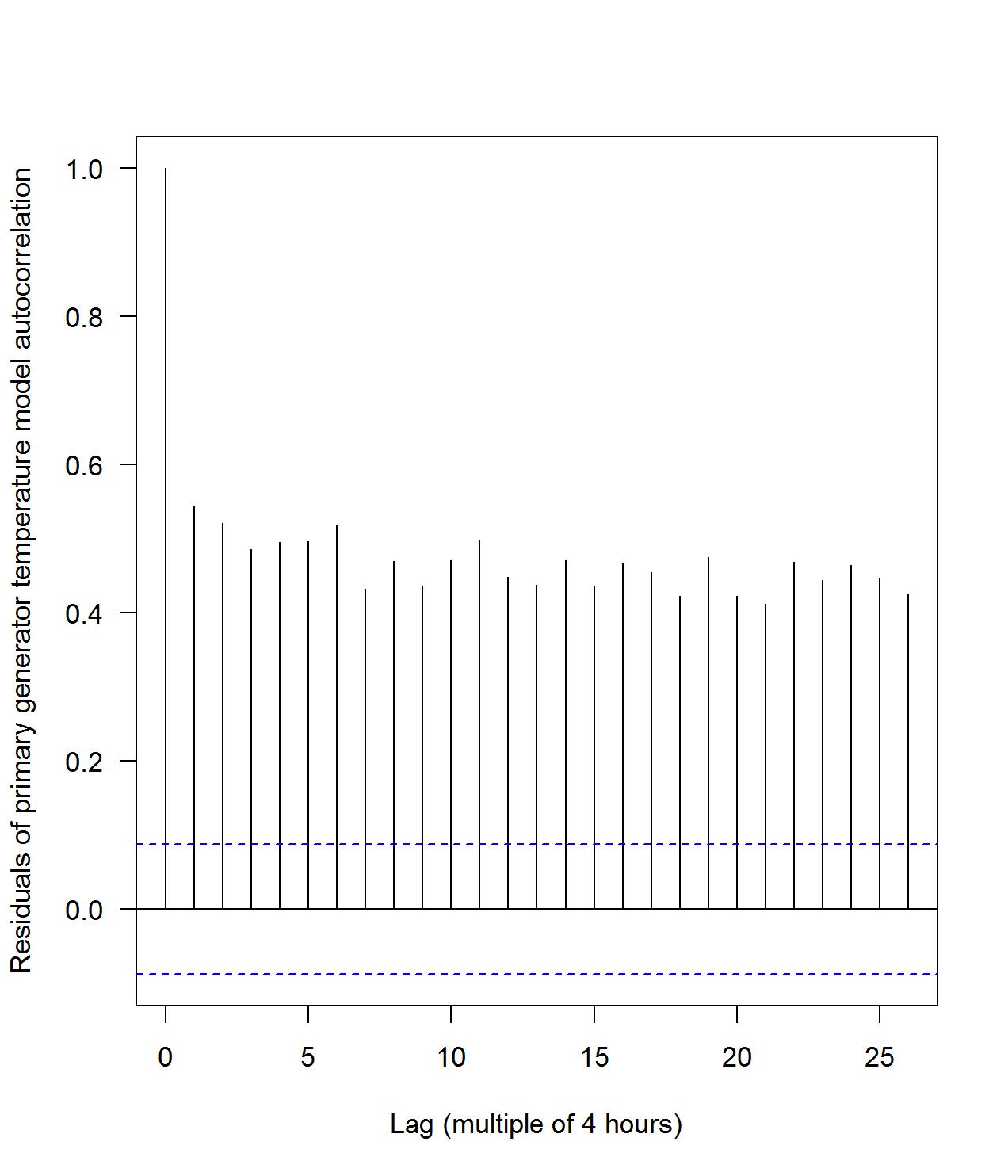}
        \caption{Autocorrelation for the residuals of the primary generator temperature model with 4 hour lag}
        \label{fig:rgentemp_aut_4h}
    \end{subfigure}
\caption{Autocorrelation plot for the primary generator temperature, using the entire data set in the left plot, and the reduced (4h-intervals) in the right plot}
\label{acf:gen1}
\end{figure}

In this section we the ideas described previously to the generator temperature. Again we consider as in-control period the  period till 12/10/2013. Next, for this period we are interested in determining the variables that are relevant for the prediction of the primary generator temperature. We consider a list of potential candidates that include among others the wind speed, the primary generator speed, the  environmental temperature, and the gearbox and bearing temperatures.
It must be noted that the power output is also a potential explanatory variable. However, due to the number of missing values, it would limit the in-control period too much and might obstruct us from being able to create a proper prediction for the generator temperature. For that reason, the power output is omitted from this model. For now, we also leave out any vibration variables, as these variables are used to produce a separate model in the next section.
Similarly to the previous model, we sub-sample the data set every 4 hours in order to reduce autocorrelation. The autocorrelation plot shows us much lower values for the 4-hour model. This is consistent with earlier observations in other models. The autocorrelation plot is depicted in Figure~\ref{acf:gen1}. Based on the 4-hour in-control data set, we  create an initial regression model, which includes all relevant explanatory variables. We then perform an all-subset selection method, which uses Mallow's $C_p$ criterion.
The coefficients and variables for the primary generator regression model are listed in Table~\ref{tab:gen1model_4h}. We see that the environmental temperature,  the generator speed and the bearing and gearbox temperatures  are included.
\begin{table}[hbt]
\centering
\begin{tabular}{|r | r|r|r|r|}
  \hline
\multicolumn{1}{|r}{} & \multicolumn{1}{r|}{Estimate} & \multicolumn{1}{r|}{Std. Error} & \multicolumn{1}{r|}{t-value} & \multicolumn{1}{r|}{p-value} \\
\hline
(Intercept) &-510.3844   &183.7866  &-2.777&  0.00648 \\
   \hline
EnvTemp      &  0.4695    & 0.1888   &2.486 &  0.01446 \\
   \hline
GenSpeed    &   0.2836     &0.1350 &  2.100&  0.03809  \\
   \hline
BrgTemp    &   -6.1955     &1.1374 & -5.447 & 3.3e-07  \\
   \hline
GbxTemp     &   8.5525  &   0.8220  &10.405  &$<$ 2e-16\\
   \hline
\end{tabular}
\caption{Regression model for the temperature of the primary generator}
\label{tab:gen1model_4h}
\end{table}

We then use this model to predict point estimates for all points in the data set, and with them compute the residuals. The residuals that correspond to the in-control period are used to compute the control limits for the Shewhart chart. The control limits of the  Shewhart chart are calculated at -23.01381 and  23.56778. In this case, we see a small amount of data which is beyond the control limits within the in-control period, and an increase of out-of-control data beyond the control limit. 3678 measurements are beyond the control limit, and in total there are 30775 data points. This means that approximately 11,95\% of the measurements is out-of-control. Moreover, the majority of these out-of-control points is accumulated in the period prior of the generator failure in December, 2014. This becomes especially evident
\begin{figure}[htb]
\centering
\includegraphics[width=0.45\textwidth]{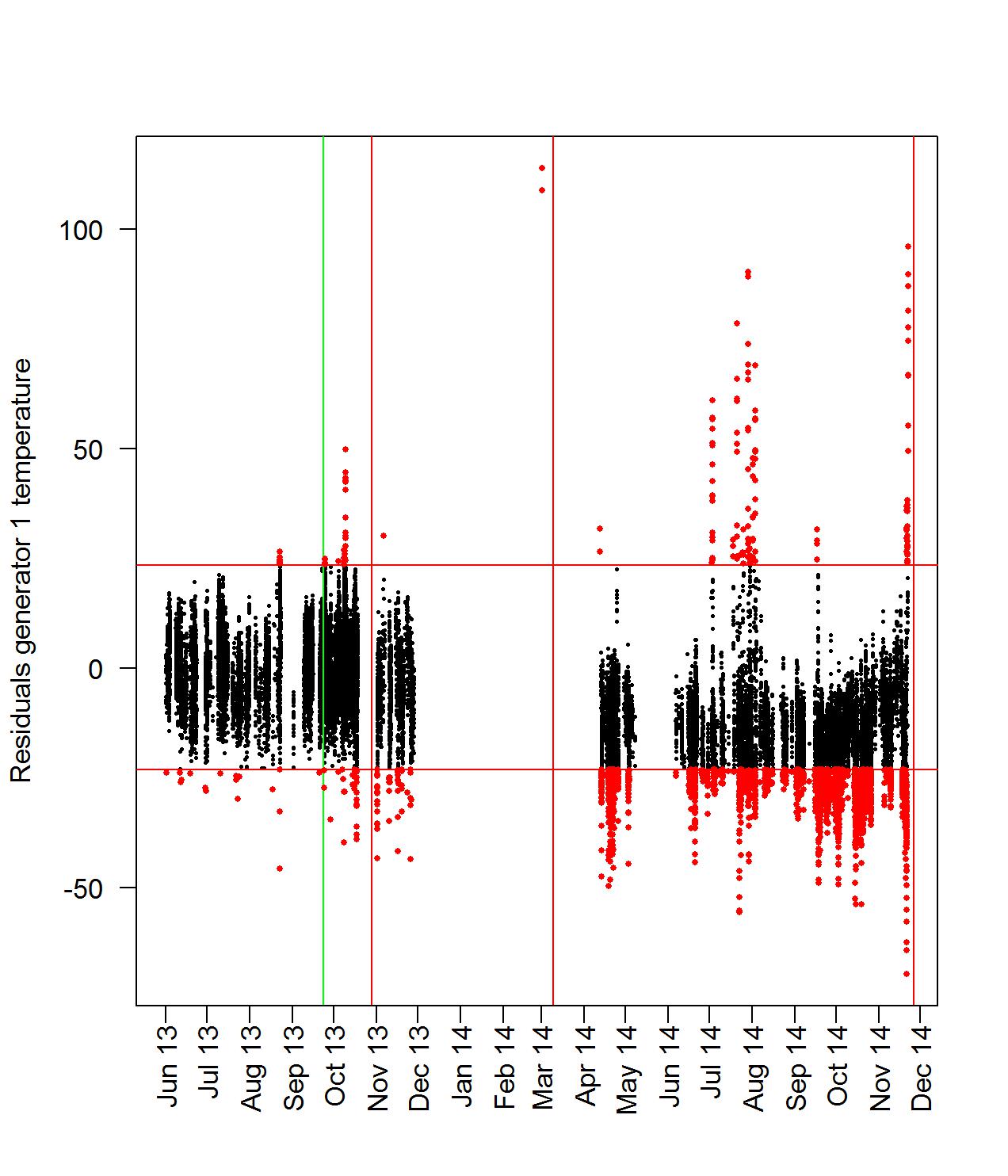}
\caption{Shewhart chart for the residuals of the primary generator temperature}
\end{figure}
when plotting the primary generator temperature versus time and color characterizing the data points which are outside the Shewhart control limits. Our result seem justifiable since there are many temperature spikes during the out-of-control period, some even up to around 130 degrees Calcium. We can interpret this as a clear indication of the imminent failure.

\begin{figure}[htb]
\centering
\includegraphics[width=0.45\textwidth]{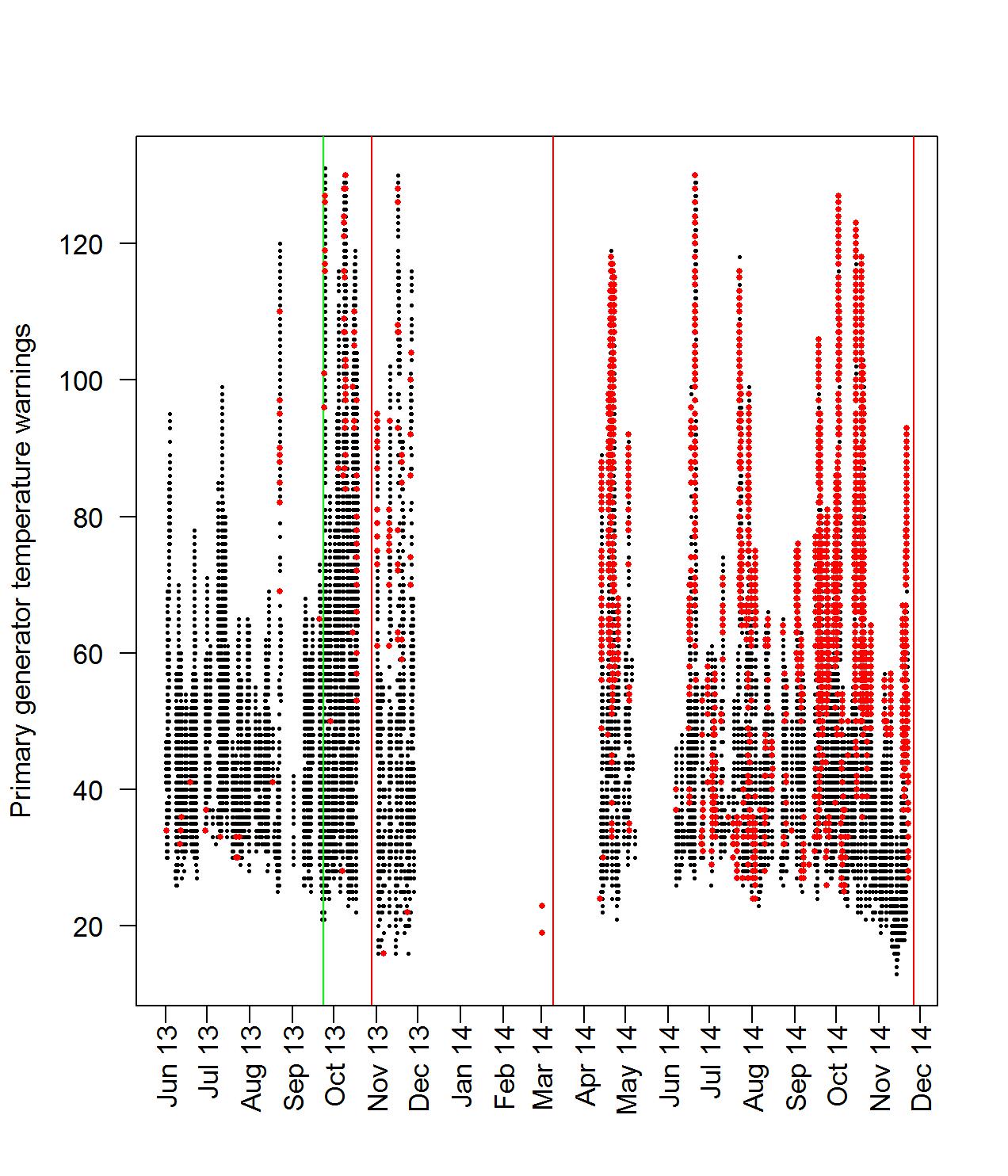}
\caption{Main generator temperature. Warnings are color-characterized in red.}
\end{figure}

It is important to note that the measured temperature of various components does not have any fixed thresholds that would indicate an out-of-control behavior, i.e. there is no clear temperature threshold that could be used to indicate an alarming behavior. However, our analysis overcomes this by
accounting for various factors including special operating conditions and investigating changes in the mean and in trend.


\subsection{Model for bearing vibrations readings}
\begin{figure}
    \centering
    \begin{subfigure}[b]{0.45\textwidth}
        \includegraphics[width=\textwidth]{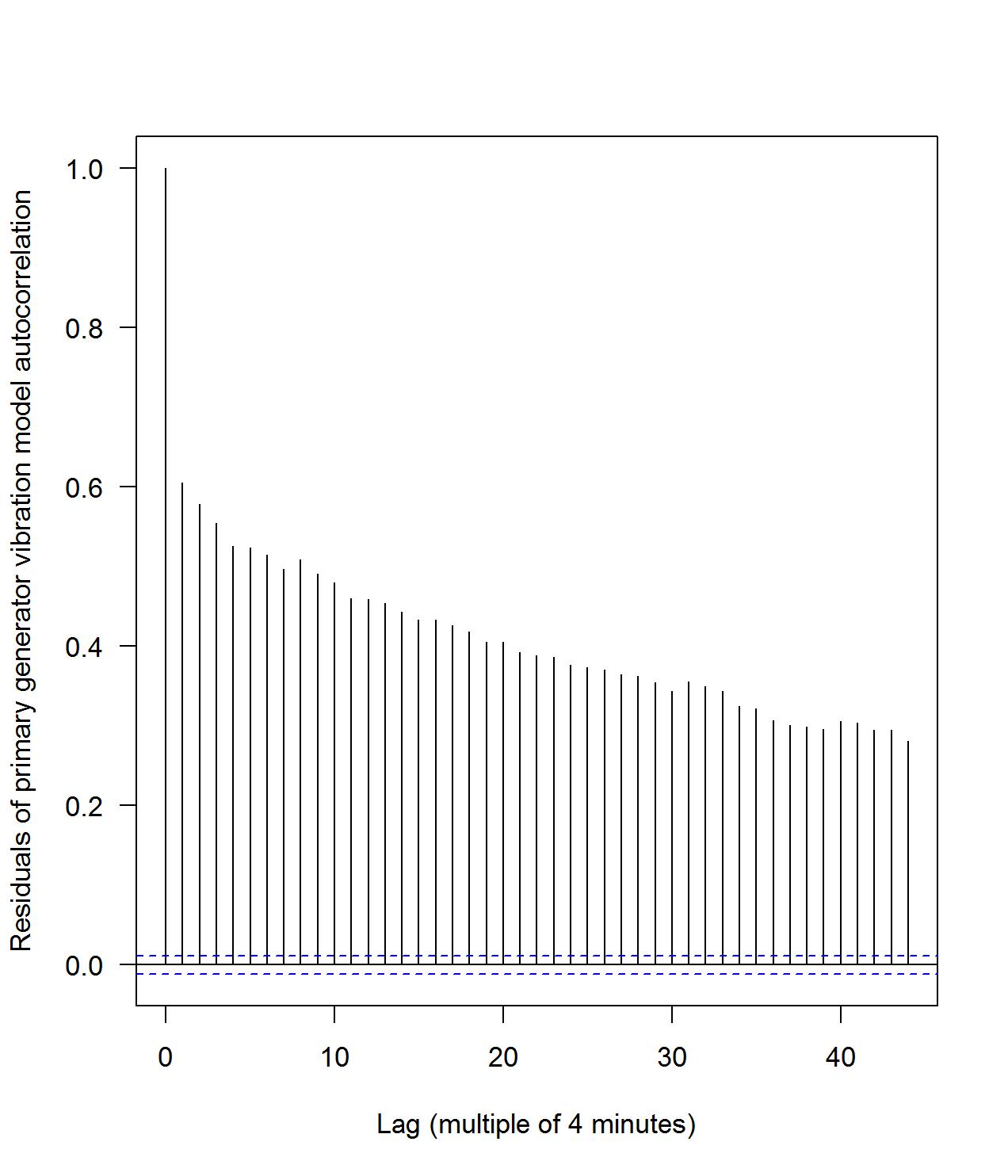}
        \caption{Autocorrelation for the residuals of the primary generator vibration model with 4 minute lag}
        \label{fig:aut_res_4m_vib}
    \end{subfigure}
    ~
   \begin{subfigure}[b]{0.45\textwidth}
        \includegraphics[width=\textwidth]{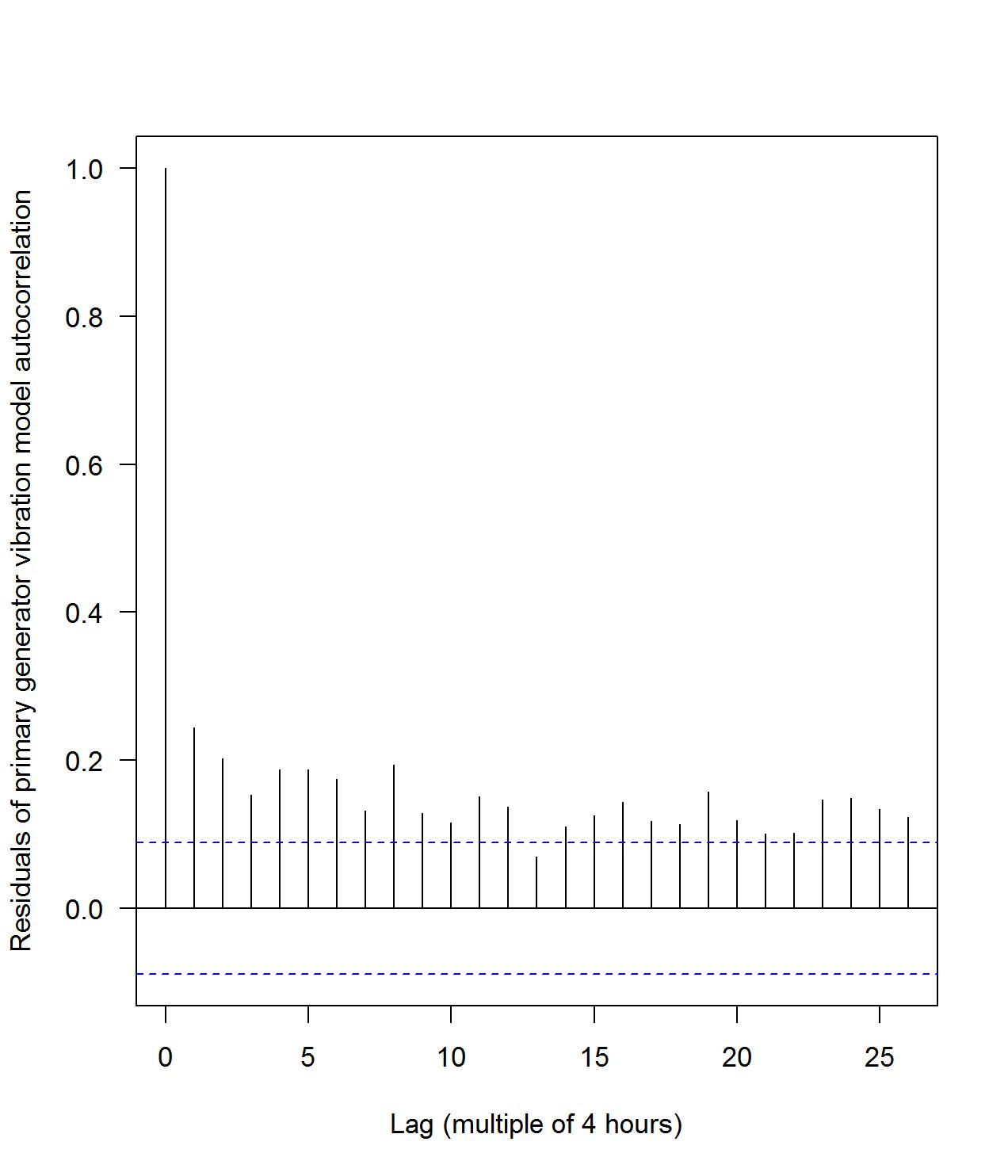}
        \caption{Autocorrelation for the residuals of the primary generator vibration model with 4 hour lag}
        \label{fig:aut_res_4h_vib}
    \end{subfigure}
\caption{Autocorrelation plot for the primary generator vibration, using the entire data set in the left plot, and the reduced (4h-intervals) in the right plot}
\label{acf:Vibgen1}
\end{figure}

Similarly to the analysis of the previous section,  we construct a regression model for the primary generator vibration readings based on the in-control 4-hour sub-sampled data set.
The autocorrelation plot  for the primary generator vibration readings is depicted in Figure~\ref{acf:Vibgen1}.
We first create an initial regression model, which include all relevant explanatory variables at various powers. We then perform an all-subset selection method, which uses Mallow's $C_p$ criterion, and choose the best model.
The coefficients and variables for the primary generator vibration regression model are listed in Table~\ref{tab:Vibgen1model_4h}. We see that the primary generator speed, wind speed and primary generator temperature are included.
\begin{table}[hbt]
\centering
\begin{tabular}{|r | r|r|r|r|}
  \hline
\multicolumn{1}{|r}{} & \multicolumn{1}{r|}{Estimate} & \multicolumn{1}{r|}{Std. Error} & \multicolumn{1}{r|}{t-value} & \multicolumn{1}{r|}{p-value} \\
\hline
(Intercept)  &  -1.299e+01&  3.166e+00&  -4.102 &8.14e-05 \\
GenSpeed   &     8.259e-03 & 2.119e-03 &  3.897 &0.000173 \\
WindSpeed${}^3$ & 1.732e-04 & 2.652e-05&   6.530 &2.47e-09 \\
Gen1Temp     &   5.691e-02&  1.278e-02  & 4.454 &2.13e-05 \\
Gen1Temp${}^2$ & -8.202e-04 & 1.941e-04 & -4.227& 5.11e-05 \\
Gen1Temp${}^3$&  3.361e-06 & 9.163e-07&   3.668 &0.000387 \\
   \hline
\end{tabular}
\caption{Regression model for the vibration of the primary generator}
\label{tab:Vibgen1model_4h}
\end{table}
We then use this model to predict point estimates for all points in the data set, and with them compute the residuals. The residuals that correspond to the in-control period are used to compute the control limits for the Shewhart chart.
The control limits of the Shewhart chart are -0.346966 and 0.3561348. In this case note that during the in-control period a few data points are outside the control limits, and this effect increases drastically as we move to the out-of-control period.

\begin{figure}[htb]
\centering
\includegraphics[width=0.45\textwidth]{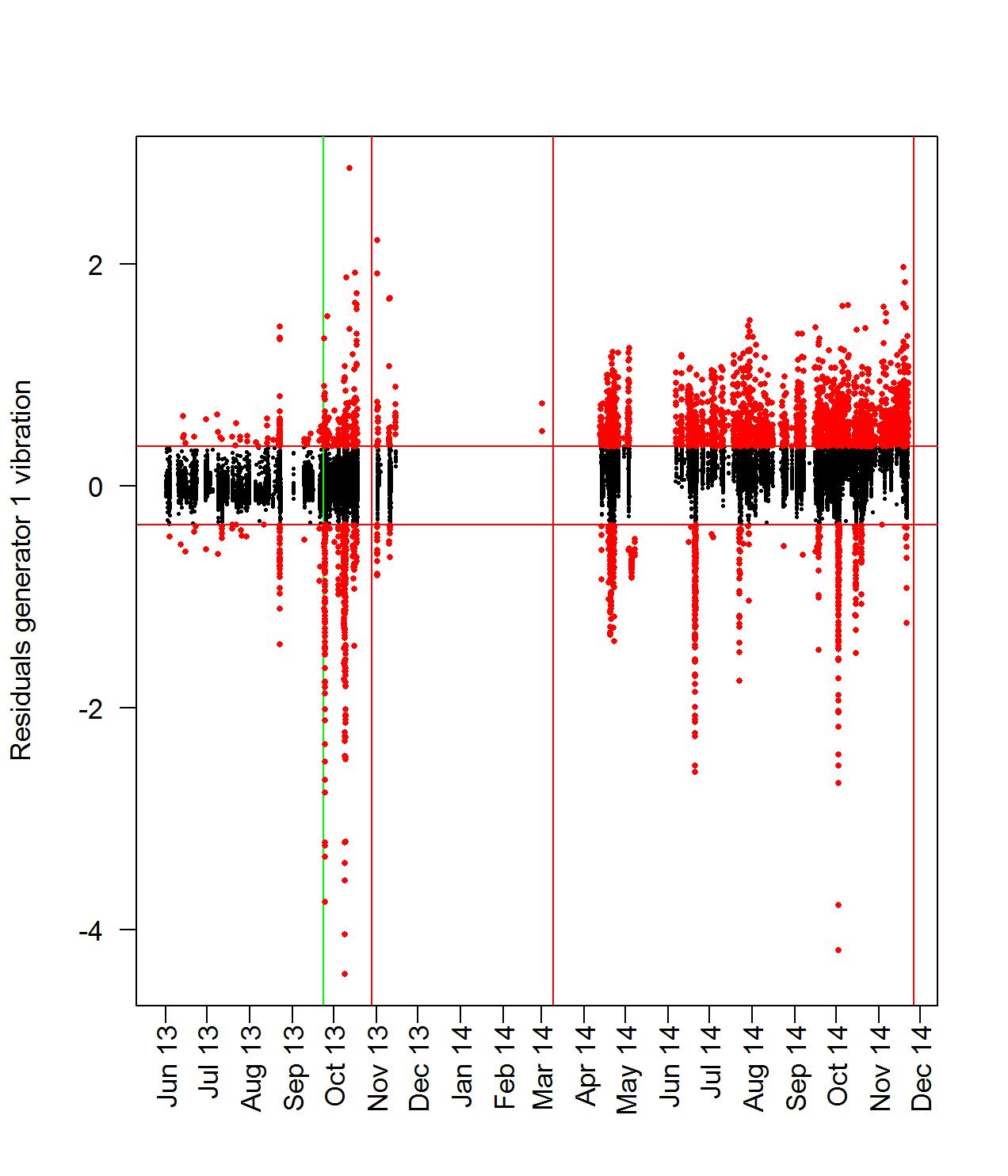}
\caption{Shewhart chart for the residuals of the primary generator vibration}
\end{figure}

For the primary generator vibration reading it is established that there are two fixed thresholds set the first at 1.06 (warning threshold) and the second set at 2.12 (alarm threshold). Based on these values, we notice that during the out-of-control period there are several readings exceeding the fixed threshold of 2.12. However, it seems that our adaptive Shewhart chart capture not only the points exceeding the fixed thresholds, but also points of very low vibrations levels that cannot be explained by the operating conditions. Furthermore, the warnings generated by the adaptive Shewhart chart are more consistent with the impression of the engineers inspecting the generator prior to the December, 2014 , failure.

\begin{figure}[htb]
\centering
\includegraphics[width=0.45\textwidth]{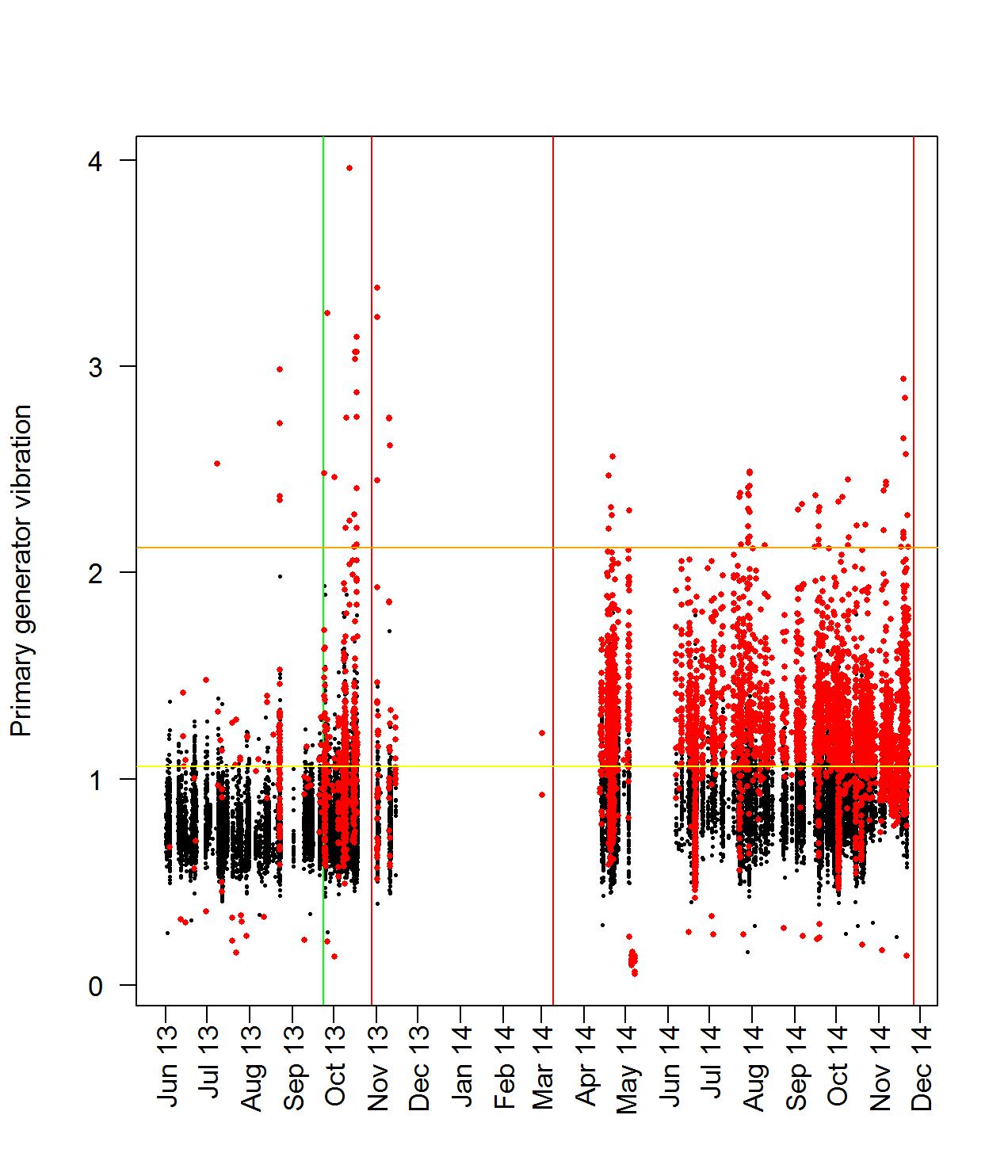}
\caption{Main generator vibration. Warnings are color-characterized in red. Fixed warning threshold at 1.06 (yellow) and fixed warning threshold at 2.12 (orange)}
\end{figure}


\section{Conclusions}\label{sec:conclusions}
In this article, we developed a statistical approach to wind turbine condition monitoring to create a system which monitors the overall health of the turbine. We identified critical relations between variables of different natures, and used these relations to construct models which are able to identify out-of-control behavior. In order to construct these models, we made use of several existing statistical methods and combined them to be suited for the application at hand. In particular, we applied Shewhart's methods for identifying the out-of-control behavior of the residuals of a regression model.
The models seem to reflect the condition of the wind turbine reasonably well.

We  are able to draw the following conclusions:
\begin{itemize}
\item Strong correlations exist between the temperature of several parts of the turbine. It is possible to create models from these variables, however, for more accurate models we need other variables as well.
\item It is possible to identify which generator is being used from the rotor speed. Each generator operates at a distinct speed.
\item We are able to apply the methods created by Shewhart to the residuals of a linear regression model so as to identify out-of-control behavior.
\item All models for the various components, except the model for the secondary generator, exhibit the same out-of-control type of behavior.
\item In the future, these models would probably be able to predict the failure of the turbine.
\end{itemize}

The case study reported in this paper has shown that condition based prognosis and diagnostics has real potential for wind turbines, and perhaps other (mechanical) applications. In order to improve this concept and improve the practical applicability, further research is suggested.
First, we suggest further research into the average run length cycle of the Shewhart charts for the residuals. Right now, a point is identified as a warning when it lies beyond certain control limits, and many warnings in a short period of time would constitute out-of-control behavior. However, the exact number of warnings that is needed in order for the behavior to be classified as out-of-control is currently unknown. Research into the average run length could provide insight into this, and help construct an objective mathematical standard on what exactly constitutes out-of-control behavior in this context, further eliminating subjectivity from the process.
Furthermore, more research could be done into the more specific identification of the root cause of failures. From the maintenance logs, we knew the primary generator malfunctioned, so it was possible to look for patterns that reflected this. If data is available from other wind turbines where other parts than the generator broke down, this might provide insight into the root causes and perhaps allow a precise identification.
On a more theoretical level, we could use more refined control charts than Shewhart charts like EWMA and CUSUM control charts. Especially self-starting CUSUM control charts  merit further research since they may be a way to avoid the difficult task of establishing  a baseline period (see \cite{HawkinsOlwell} and \cite{HawkinsQiuKang}).

\section*{Acknowledgments}
The work of the authors is supported by a TKI Project, DAISY4OFFSHORE.  In addition, the work of Stella Kapodistria in the begging of this project was supported by a Dinalog project, CAMPI, and is currently supported by an NWO Gravitation Project, NETWORKS. The authors would like to thank the companies and institutes involved in the DAISY4OFFSHORE consortium for their time and advice in the preparation of this work.

\end{document}